# Non-equilibrium effects on electron-phonon coupling constant in metals

Wuli Miao, Moran Wang[†]

*Key Laboratory for Thermal Science and Power Engineering of Ministry of Education, Department of Engineering Mechanics and CNMM, Tsinghua University, Beijing 100084, China*

## Abstract

Understanding of the energy exchange between electrons and phonons in metals is important for micro- and nano-manufacturing and system design. The electron-phonon (e-ph) coupling constant is to describe such exchange strength, yet its variation remains still unclear at micro- and nanoscale where the non-equilibrium effects are significant. In this work, an e-ph coupling model is proposed by transforming the full scattering terms into relaxation time approximation forms in the coupled electron and phonon Boltzmann transport equations. Consequently, the non-equilibrium effects are included in the calculation of e-ph coupling constant. The coupling model is verified by modeling the ultrafast dynamics in femtosecond pump-probe experiments on metal surface, which shows consistent results with the full integral treatment of scattering terms. The e-ph coupling constant is strongly reduced due to both the temporal non-equilibrium between different phonon branches and the spatial non-equilibrium of electrons in confined space. The present work will promote not only a fundamental understanding of the e-ph coupling constant but also the theoretical description of coupled electron and phonon transport at micro- and nanoscale.

---

[†] Corresponding author; Tel: 86-10-627-87498; Email: mrwang@tsinghua.edu.cn (MW)

## 1. Introduction

Excitation of a metal by a femtosecond pulse laser involves a very complex process including electron-photon, electron-electron, electron-phonon, phonon-phonon interactions [1, 2]. In spite of its complexity, the pre-assumption of thermal equilibrium for electrons and phonons subsystems respectively is usually adopted, and electrons will exchange energy with phonons through electron-phonon (e-ph) scattering [3, 4]. However, extensive experimental and theoretical studies have demonstrated the failure of the thermal equilibrium assumption due to both the non-thermalized electrons and the non-equilibrium between different phonon polarizations [5-12]. Besides, when the characteristic length of metals shrinks to nanoscale in micro- and nano-electronics, the spatial non-equilibrium effect arising from size effect becomes significant [13, 14]. These non-equilibrium effects will further influence the energy exchange between electrons and phonons.

In order to describe the strength of energy exchange in metals, the e-ph coupling constant defined as energy transfer rate per unit volume and per temperature difference between electrons and phonons is quantitatively introduced [15]. In 1957, Kaganov *et al* studied the energy transfer through e-ph coupling firstly based on the Boltzmann transport theory and formulated an expression of the e-ph coupling constant. This expression connected the e-ph coupling constant to an empirical electron relaxation time, which was based on both the free electron gas model and thermal equilibrium for electrons and phonons subsystems respectively [16]. After nearly 30 years, Allen insightfully related the coupling function to the Eliashberg function in superconductivity, and then derived the corresponding theoretic formula of the e-ph coupling constant [17]. Later on, Allen's formula was validated by the femtosecond pulse laser experiments [18] and became a basic principle for most metals. However, the key assumption behind Allen's derivation was that both electrons and phonons were in thermal quasi-equilibrium corresponding to their respective temperatures, which might be not valid at micro- and nanoscale [13, 14].

Generally, the non-equilibrium effects are significant when the characteristic length of the system is comparable to the mean free path of heat carriers or the characteristic time of the process is comparable to the relaxation time. As a result, the transport properties like thermal conductivity and electronic conductivity decrease notably from the standard bulk values [19, 20]. In terms of the e-ph coupling constant, the results are not clear for thin films and nanoparticles in the literature by the different researchers using the femtosecond pump-probe experiments [21-28]. It remains inconclusive how the e-ph coupling constant varies when the characteristic length of metals reduces to nanoscale size. In ultra-fast dynamics irradiated by a femtosecond pulse laser, non-thermalized electrons exist due to the finite electron-electron relaxation time [5-8, 12]. It means that electrons have transferred energy to phonons before the electron sub-system reaches thermal equilibrium. An experiment in a pulse-heated metal showed that non-thermalized electrons make e-ph energy relaxation process slower [7]. The theoretical analysis also showed that the e-ph coupling rate is weakened comparing to the thermalized limit at low excitation intensities [2, 29]. Moreover, the coupling strength between electrons and different phonon branches is often different and thus the non-equilibrium between different phonon polarizations occurs [10, 11]. However, it remains to investigate how the e-ph coupling constant is influenced by this temporal non-equilibrium effect between different phonon branches. Therefore, this work aims to reveal the variation of e-ph coupling constant in metals at micro- and nanoscale when the spatial or temporal non-equilibrium effects are significant.

Current calculations of the e-ph coupling constant are generally based on the semi-classical e-ph scattering term in Boltzmann transport equation which includes an e-ph scattering matrix element representing alterations of electron states by absorbing or emitting a phonon [30]. The prevailing method is to solve the e-ph scattering matrix element using *ab initio* calculation and thus the e-ph coupling constant is computed by summing the energy change in e-ph scattering process [31-33]. This commonly used method is also based on the assumption of equilibrium for electrons and phonons subsystems respectively. In contrast, the direct solution of the coupled electron and phonon Boltzmann transport equations (BTEs) provides an alternative choice. Based on the evolution of distribution functions, the non-equilibrium effects can be intrinsically considered. However, the challenge of the coupled electron and phonon BTEs lies on treatment of the scattering term due to its integral form and coupling between electron and phonon distribution functions, which makes the direct solution very difficult [2, 9, 29, 34]. In addition to the linear response assumption [9], the integral form of scattering terms is solvable as done by Rethfeld and Ono, when neglecting the drift term [2, 34]. However, this practice is hard to consider the drift term simultaneously which is necessary for the transport issues. Besides, there are also some macroscopic models to capture the coupled electron and phonon transport, including two-temperature model (TTM) and multi-temperature model (MTM) with the e-ph coupling constant as an input parameter [10, 15, 35-37]. Therefore, the other aim of this study is to propose a feasible treatment of scattering term in the coupled electron and phonon BTEs. For solution of BTEs, two categories of numerical schemes are currently available, including the stochastic method like Monte Carlo (MC) scheme [38, 39] and the deterministic method such as discrete-ordinate-method (DOM) [40, 41]. The treatment of e-ph coupling in MC schemes is still challenging so that we choose the DOM scheme first. Last, we want to clarify the term "non-equilibrium" in the present work for a better understanding, which contains two kinds of meaning as below: (i) the "overall" non-equilibrium as the temperature of electrons is different with that of phonons; (ii) the "local" non-equilibrium as electron and phonon each subsystem is not equilibrium. When dealing with the e-ph coupling constant, we are based on the "overall" non-equilibrium. It is what the theoretical formula and *ab initio* calculation do without consideration of the "local" non-equilibrium. In comparison, the present work also takes the "local" non-equilibrium into account and investigates the influence of this "local" non-equilibrium on the e-ph coupling constant.

The remaining of this article is organized as below. In Section 2, the theoretical derivation of the scattering model for the coupled electron and phonon BTEs is provided and the computational scheme for numerical solution is presented. The proposed coupling model is verified by modeling the ultra-fast dynamics process in femtosecond pump-probe experiments. The variation of the e-ph coupling constant is studied in Section 3 when non-equilibrium between different phonon branches or non-equilibrium of electron exists. The concluding remarks are finally made in Section 4.

## 2. Theoretical model and numerical method

### 2.1 Coupled electron and phonon BTEs under relaxation time approximation

The coupled electron and phonon BTEs without magnetic field are expressed as [42]:

$$\frac{\partial f_{\mathbf{k}}}{\partial t} + \mathbf{v}_e \cdot \nabla_{\mathbf{r}} f_{\mathbf{k}} - \frac{e\mathbf{E}}{\hbar} \cdot \nabla_{\mathbf{k}} f_{\mathbf{k}} = \Omega_{e-ph}, \tag{1}$$

$$\frac{\partial n_{\mathbf{Q,p}}}{\partial t} + \mathbf{v}_{ph,p} \cdot \nabla_{\mathbf{r}} n_{\mathbf{Q,p}} = \Omega_{ph-e} + \Omega_{ph-ph}, \tag{2}$$

where $f_{\mathbf{k}} \equiv f(\mathbf{r}, \mathbf{k}, t)$ is the electron distribution function denoting the electron occupation number around the wave vector $\mathbf{k}$ and the spatial position $\mathbf{r}$ at the moment $t$ and $n_{\mathbf{Q,p}} \equiv n(\mathbf{Q}, \mathrm{p}, \mathbf{r}, t)$ is the phonon distribution function with p the phonon polarizations, $\mathbf{Q}$ the phonon wave vector. $\mathbf{v}_e$ is the electron drift velocity with $\mathbf{v}_{ph,p}$ the phonon group velocity of different polarizations. $\hbar$ is the reduced Planck constant and $\mathbf{E}$ is the effective electric field with $e$ the element charge. In the scattering process, the e-ph collision dominates for the alteration of electron distribution function, which is represented by $\Omega_{e-ph}$. Under low-fluence excitation and perturbation, the electron-electron interaction is very weak due to the greatly restricted scattering phase space by Pauli exclusion principle [7, 8]. Considering also the screened effect, the contribution from this electron-electron interaction is neglected as a first step [20, 42]. $\Omega_{ph-e}$ and $\Omega_{ph-ph}$ denote the alteration of phonon distribution function by the phonon-electron (ph-e) scattering and phonon-phonon (ph-ph) scattering respectively. The imperfection scattering is not considered as a first step, which can be incorporated in a straightforward way in the near future. Based on the Fermi's golden rule, the integral forms of $\Omega_{e-ph}$ and $\Omega_{ph-e}$ are formulated as:

$$\Omega_{e-ph} = -\frac{2\pi}{\hbar}\sum_{\mathbf{Q},\mathrm{p}}\left|g\left(\mathbf{k}',\mathbf{k},\mathrm{p}\right)\right|^2\left\{f_{\mathbf{k}}\left(1-f_{\mathbf{k}'}\right)\left[\left(n_{\mathbf{Q},\mathrm{p}}+1\right)\delta\left(\varepsilon_{\mathbf{k}}-\varepsilon_{\mathbf{k}'}-\hbar\omega_{\mathbf{Q},\mathrm{p}}\right)+n_{\mathbf{Q},\mathrm{p}}\delta\left(\varepsilon_{\mathbf{k}}-\varepsilon_{\mathbf{k}'}+\hbar\omega_{\mathbf{Q},\mathrm{p}}\right)\right]\right.$$
$$\left.-\left(1-f_{\mathbf{k}}\right)f_{\mathbf{k}'}\left[\left(n_{\mathbf{Q},\mathrm{p}}+1\right)\delta\left(\varepsilon_{\mathbf{k}}-\varepsilon_{\mathbf{k}'}+\hbar\omega_{\mathbf{Q},\mathrm{p}}\right)+n_{\mathbf{Q},\mathrm{p}}\delta\left(\varepsilon_{\mathbf{k}}-\varepsilon_{\mathbf{k}'}-\hbar\omega_{\mathbf{Q},\mathrm{p}}\right)\right]\right\}$$

$$(3)$$

$$\Omega_{ph-e} = -\frac{2\pi}{\hbar}\sum_{\mathbf{k}}\left|g\left(\mathbf{k}',\mathbf{k},\mathrm{p}\right)\right|^2 f_{\mathbf{k}}\left(1-f_{\mathbf{k}'}\right)\left[n_{\mathbf{Q},\mathrm{p}}\delta\left(\varepsilon_{\mathbf{k}}-\varepsilon_{\mathbf{k}'}+\hbar\omega_{\mathbf{Q},\mathrm{p}}\right)-\left(n_{\mathbf{Q},\mathrm{p}}+1\right)\delta\left(\varepsilon_{\mathbf{k}}-\varepsilon_{\mathbf{k}'}-\hbar\omega_{\mathbf{Q},\mathrm{p}}\right)\right],(4)$$

where $g\left(\mathbf{k}',\mathbf{k},\mathrm{p}\right)$ is the e-ph scattering matrix element with its square representing the probability of electron transition from the state $\mathbf{k}$ with energy $\varepsilon_{\mathbf{k}}$ to the state $\mathbf{k}'$ with energy $\varepsilon_{\mathbf{k}'}$ by absorbing or emitting a phonon with frequency $\omega_{\mathbf{Q},\mathrm{p}}$ [19, 42]. The summation of electron wave vector state includes the spin degeneracy. Especially, the electron distribution function and phonon distribution function are strongly coupled with each other through the collision terms. Besides, the BTEs are integro-differential equations, which makes the direct solution very challenging.

Within the linear response regime, the electron and phonon distribution function can be written as $f_{\mathbf{k}} = f_{\mathbf{k}}^{eq}\left(\tilde{T}_e\right)+\Delta f_{\mathbf{k}}$, $n_{\mathbf{Q},\mathrm{p}} = n_{\mathbf{Q},\mathrm{p}}^{eq}\left(\tilde{T}_{ph}\right)+\Delta n_{\mathbf{Q},\mathrm{p}}$ respectively with the deviation part a small quantity $\Delta f_{\mathbf{k}} \ll f_{\mathbf{k}}^{eq}\left(\tilde{T}_e\right)$, $\Delta n_{\mathbf{Q},\mathrm{p}} \ll n_{\mathbf{Q},\mathrm{p}}^{eq}\left(\tilde{T}_{ph}\right)$ from the equilibrium state $f_{\mathbf{k}}^{eq}\left(\tilde{T}_e\right)$, $n_{\mathbf{Q},\mathrm{p}}^{eq}\left(\tilde{T}_{ph}\right)$ at the pseudo-temperature $\tilde{T}_e$, $\tilde{T}_{ph}$ separately. In this way, the e-ph scattering term in Eq. (3) and the ph-e scattering term in Eq. (4) are reformulated into

$$\Omega_{e-ph} = -\frac{2\pi}{\hbar} \sum_{\mathbf{Q},p} \left| g\left(\mathbf{k'},\mathbf{k},p\right) \right|^2 \left\{ \left[ \left( f_{\mathbf{k}}^{eq}(\tilde{T}_e) - f_{\mathbf{k'}}^{eq}(\tilde{T}_e) \right) \left( n_{\mathbf{Q},p} - n_{\mathbf{Q},p}^{eq}(\tilde{T}_e) \right) \right. \right.$$
$$+ \Delta f_{\mathbf{k}} \left( n_{\mathbf{Q},p}^{eq}(\tilde{T}_{ph}) + 1 - f_{\mathbf{k'}}^{eq}(\tilde{T}_e) \right)$$
$$\left. - \Delta f_{\mathbf{k'}} \left( n_{\mathbf{Q},p}^{eq}(\tilde{T}_{ph}) + f_{\mathbf{k}}^{eq}(\tilde{T}_e) \right) \right] \delta \left( \varepsilon_{\mathbf{k}} - \varepsilon_{\mathbf{k'}} - \hbar \omega_{\mathbf{Q},p} \right)$$
$$+ \left[ \left( f_{\mathbf{k}}^{eq}(\tilde{T}_e) - f_{\mathbf{k'}}^{eq}(\tilde{T}_e) \right) \left( n_{\mathbf{Q},p} - n_{\mathbf{Q},p}^{eq}(\tilde{T}_e) \right) \right.$$
$$+ \Delta f_{\mathbf{k}} \left( n_{\mathbf{Q},p}^{eq}(\tilde{T}_{ph}) + f_{\mathbf{k'}}^{eq}(\tilde{T}_e) \right)$$
$$\left. \left. - \Delta f_{\mathbf{k'}} \left( n_{\mathbf{Q},p}^{eq}(\tilde{T}_{ph}) + 1 - f_{\mathbf{k}}^{eq}(\tilde{T}_e) \right) \right] \delta \left( \varepsilon_{\mathbf{k}} - \varepsilon_{\mathbf{k'}} + \hbar \omega_{\mathbf{Q},p} \right) \right\} \quad , \ (5)$$

$$\Omega_{ph-e} = -\frac{2\pi}{\hbar} \sum_{\mathbf{k}} \left| g\left(\mathbf{k'},\mathbf{k},p\right) \right|^2 \left\{ \left( f_{\mathbf{k}}^{eq}(\tilde{T}_e) - f_{\mathbf{k'}}^{eq}(\tilde{T}_e) \right) \left( n_{\mathbf{Q},p} - n_{\mathbf{Q},p}^{eq}(\tilde{T}_e) \right) \right.$$
$$+ \Delta f_{\mathbf{k}} \left( n_{\mathbf{Q},p}^{eq}(\tilde{T}_{ph}) + f_{\mathbf{k'}}^{eq}(\tilde{T}_e) \right)$$
$$\left. - \Delta f_{\mathbf{k'}} \left( n_{\mathbf{Q},p}^{eq}(\tilde{T}_{ph}) + 1 - f_{\mathbf{k}}^{eq}(\tilde{T}_e) \right) \right\} \delta \left( \varepsilon_{\mathbf{k}} - \varepsilon_{\mathbf{k'}} + \hbar \omega_{\mathbf{Q},p} \right) \quad . \ (6)$$

Generally, the pseudo-temperature of electron, $\tilde{T}_e$, is different from that of phonon, $\tilde{T}_{ph}$ such that the expression $n_{\mathbf{Q},p} - n_{\mathbf{Q},p}^{eq}(\tilde{T}_e)$ is formulated with $n_{\mathbf{Q},p}^{eq}(\tilde{T}_e)$ at the electron pseudo-temperature. This expression contains the non-equilibrium phonon part $\Delta n_{\mathbf{Q},p}$ and the difference between the electron and phonon pseudo-temperature $n_{\mathbf{Q},p}^{eq}(\tilde{T}_{ph}) - n_{\mathbf{Q},p}^{eq}(\tilde{T}_e)$.

The non-equilibrium phonon in electron scattering term in Eq. (5) is called phonon drag [43]. Correspondingly, the non-equilibrium electron in phonon scattering term as shown by $\Delta f_{\mathbf{k}}$ and $\Delta f_{\mathbf{k'}}$ in Eq. (6) is called electron drag. These are mutual effects which fully couple electron and phonon BTEs [44]. Nevertheless, the drag effects play a non-negligible role at low temperature as the ph-ph scattering are greatly weakened and phonons cannot be back to local equilibrium quickly. At the temperature scope concerned in the present work, the drag effects are neglected as a first step. Besides, when dealing with the e-ph scattering term in Eq. (5), the remaining terms $n_{\mathbf{Q},p}^{eq}(\tilde{T}_{ph}) - n_{\mathbf{Q},p}^{eq}(\tilde{T}_e)$ implicitly contained in $n_{\mathbf{Q},p} - n_{\mathbf{Q},p}^{eq}(\tilde{T}_e)$ are almost cancelled with each other through the first-order Taylor expansion of the term $f_{\mathbf{k}}^{eq}(\tilde{T}_e) - f_{\mathbf{k}'}^{eq}(\tilde{T}_e)$. In other words, the terms $n_{\mathbf{Q},p} - n_{\mathbf{Q},p}^{eq}(\tilde{T}_e)$ in electron scattering term are negligible due to the inappreciable drag effect at the evaluated temperature and the nearly cancelling summation of absorption and emission process. Furthermore, under the isotropic scattering picture, the non-equilibrium $\Delta f_{\mathbf{k}'}$ of other electron states in Eq. (5) vanishes by integration [45]. Therefore, the e-ph scattering term in Eq. (5) and the ph-e scattering term in Eq. (6) are simplified into the form of relaxation time approximation:

$$\Omega_{e-ph} = -\frac{2\pi}{\hbar} \sum_{\mathbf{Q},p} \left| g\left(\mathbf{k}',\mathbf{k},p\right) \right|^2 \left\{ \Delta f_{\mathbf{k}} \left( n_{\mathbf{Q},p}^{eq}(\tilde{T}_{ph}) + 1 - f_{\mathbf{k}'}^{eq}(\tilde{T}_e) \right) \delta\left( \varepsilon_{\mathbf{k}} - \varepsilon_{\mathbf{k}'} - \hbar\omega_{\mathbf{Q},p} \right) \right.$$
$$\left. + \Delta f_{\mathbf{k}} \left( n_{\mathbf{Q},p}^{eq}(\tilde{T}_{ph}) + f_{\mathbf{k}'}^{eq}(\tilde{T}_e) \right) \delta\left( \varepsilon_{\mathbf{k}} - \varepsilon_{\mathbf{k}'} + \hbar\omega_{\mathbf{Q},p} \right) \right\} \quad , \quad (7)$$
$$= -\frac{\Delta f_{\mathbf{k}}}{\tau_{\mathbf{k},e-ph}} = -\frac{f_{\mathbf{k}} - f_{\mathbf{k}}^{eq}(\tilde{T}_e)}{\tau_{\mathbf{k},e-ph}}$$

$$\Omega_{ph-e} = -\frac{2\pi}{\hbar} \sum_{\mathbf{k}} \left| g\left(\mathbf{k}',\mathbf{k},p\right) \right|^2 \left( f_{\mathbf{k}}^{eq}(\tilde{T}_e) - f_{\mathbf{k}'}^{eq}(\tilde{T}_e) \right) \left( n_{\mathbf{Q},p} - n_{\mathbf{Q},p}^{eq}(\tilde{T}_e) \right) \delta\left( \varepsilon_{\mathbf{k}} - \varepsilon_{\mathbf{k}'} + \hbar\omega_{\mathbf{Q},p} \right) \quad ,$$
$$= -\frac{n_{\mathbf{Q},p} - n_{\mathbf{Q},p}^{eq}(\tilde{T}_e)}{\tau_{\mathbf{Q},p,ph-e}}$$

$$(8)$$

with the e-ph and ph-e scattering relaxation time defined separately as:

$$\frac{1}{\tau_{\mathbf{k},e-ph}} = \frac{2\pi}{\hbar} \sum_{\mathbf{Q},p} \left| g\left(\mathbf{k}',\mathbf{k},p\right) \right|^2 \left\{ \left( n_{\mathbf{Q},p}^{eq}(\tilde{T}_{ph}) + 1 - f_{\mathbf{k}'}^{eq}(\tilde{T}_e) \right) \delta\left( \varepsilon_{\mathbf{k}} - \varepsilon_{\mathbf{k}'} - \hbar\omega_{\mathbf{Q},p} \right) \right.$$
$$\left. + \left( n_{\mathbf{Q},p}^{eq}(\tilde{T}_{ph}) + f_{\mathbf{k}'}^{eq}(\tilde{T}_e) \right) \delta\left( \varepsilon_{\mathbf{k}} - \varepsilon_{\mathbf{k}'} + \hbar\omega_{\mathbf{Q},p} \right) \right\} \quad , \quad (9)$$

$$\frac{1}{\tau_{\mathbf{Q},p,ph-e}} = \frac{2\pi}{\hbar} \sum_{\mathbf{k}} \left| g\left(\mathbf{k}',\mathbf{k},p\right) \right|^2 \left( f_{\mathbf{k}}^{eq}(\tilde{T}_e) - f_{\mathbf{k}'}^{eq}(\tilde{T}_e) \right) \delta\left( \varepsilon_{\mathbf{k}} - \varepsilon_{\mathbf{k}'} + \hbar\omega_{\mathbf{Q},p} \right). \quad (10)$$

The assumption of isotropic electron-acoustic phonon scattering in metals is nearly valid at temperature higher than Debye temperature.

Subsequently, we adopt the relaxation time approximation for three-phonon

Umklapp scattering process with the normal scattering and high-order phonon-phonon scattering negligible at the phonon temperature regime in the present work. Thus, the ph-ph scattering is formulated as:

$$\Omega_{ph-ph} = -\frac{n_{\mathbf{Q},p} - n_{\mathbf{Q},p}^{eq}(\tilde{T}_{ph})}{\tau_{U,ph-ph}}, \tag{11}$$

with the Umklapp scattering spectral relaxation time defined as [46]

$$\frac{1}{\tau_{U,ph-ph}(\omega,p,\tilde{T}_{ph})} = B_U \omega^2 \tilde{T}_{ph} \exp\left(-\Theta_p/3\tilde{T}_{ph}\right). \tag{12}$$

In Eq. (12), $B_U = \dfrac{\hbar \gamma_p^2}{M \Theta_p v_{ph,p}^2}$ with the Grüneisen parameter $\gamma_p$ and Debye temperature $\Theta_p$ for different phonon polarizations, $M$ being the average atomic mass.

The assumption of distinct pseudo-temperatures for TA and LA phonons when dealing with the e-ph scattering term in Eq. (5) and ph-ph scattering term in Eq. (11) seems to be more consistent with the present study of non-equilibrium. However, the identical pseudo-temperature for TA and LA phonons is adopted due to the following two aspects: (i) the influence of phonon pseudo-temperature on the e-ph scattering term is only reflected in the expression of relaxation time in Eq. (9), where the distinguishing treatment of TA and LA phonons has negligible impact; (ii) ph-ph scattering pushes TA and LA phonons towards equilibrium and thus one representative phonon pseudo-temperature in ph-ph scattering term is usually introduced [47]. Therefore, we assume an overall phonon pseudo-temperature in dealing with e-ph and ph-ph scattering term as a first step.

For thermal transport, the contribution from the built-in electric field is negligibly small in metals [20, 48] and neglected as demonstrated in our previous work [38]. With also the assumption of isotropic band structure and dispersion relation, the relaxation

times of e-ph scattering and ph-e scattering are averaged on the same energy state as

$$\frac{1}{\tau_{e-ph}(\varepsilon)} = \frac{1}{D_e(\varepsilon)} \sum_{\mathbf{k}} \frac{1}{\tau_{\mathbf{k},e-ph}} \delta(\varepsilon - \varepsilon_{\mathbf{k}}),$$  (13)

$$\frac{1}{\tau_{ph-e}(\omega,\mathbf{p})} = \frac{1}{D_{ph}(\omega,\mathbf{p})} \sum_{\mathbf{Q}} \frac{1}{\tau_{\mathbf{Q},\mathbf{p},ph-e}} \delta(\omega - \omega_{\mathbf{Q},\mathbf{p}}).$$  (14)

In Eqs. (13) and (14), $D_e(\varepsilon)$ and $D_{ph}(\omega,\mathbf{p})$ are the density of states (DOS) for

electron including the spin degeneracy and for phonon at different polarizations,

respectively. With the help of the property of the Dirac $\delta$ function, the following

coupling functions are introduced as [34]

$$C_{e-ph}(\varepsilon,\varepsilon',\omega,\mathbf{p}) = \frac{1}{\hbar D_e(\varepsilon)} \sum_{\mathbf{k},\mathbf{Q}} |g(\mathbf{k}',\mathbf{k},\mathbf{p})|^2 \delta(\varepsilon - \varepsilon_{\mathbf{k}}) \delta(\varepsilon' - \varepsilon_{\mathbf{k}'}) \delta(\omega - \omega_{\mathbf{Q},\mathbf{p}}),$$  (15)

$$C_{ph-e}(\varepsilon,\varepsilon',\omega,\mathbf{p}) = \frac{1}{\hbar D_{ph}(\omega,\mathbf{p})} \sum_{\mathbf{k},\mathbf{Q}} |g(\mathbf{k}',\mathbf{k},\mathbf{p})|^2 \delta(\varepsilon - \varepsilon_{\mathbf{k}}) \delta(\varepsilon' - \varepsilon_{\mathbf{k}'}) \delta(\omega - \omega_{\mathbf{Q},\mathbf{p}}).$$  (16)

with the equation $C_{ph-e}(\varepsilon,\varepsilon',\omega,\mathbf{p}) = \dfrac{D_e(\varepsilon)}{D_{ph}(\omega,\mathbf{p})} C_{e-ph}(\varepsilon,\varepsilon',\omega,\mathbf{p})$ . The coupling

function in Eq. (15) can be determined by *ab initio* calculation, or related to the

Eliashberg function when neglecting the energy dependence of electron states [17, 34].

Thus the spectral relaxation times in Eq. (13) and (14) are formulated respectively as:

$$\frac{1}{\tau_{e-ph}(\varepsilon)} = 2\pi \sum_{\mathbf{p}} \int \left\{ \left[ n_\omega^{eq}(\tilde{T}_{ph}) + 1 - f_{\varepsilon-\hbar\omega}^{eq}(\tilde{T}_e) \right] C_{e-ph}(\varepsilon,\varepsilon-\hbar\omega,\omega,\mathbf{p}) \right.$$
$$\left. + \left[ n_\omega^{eq}(\tilde{T}_{ph}) + f_{\varepsilon+\hbar\omega}^{eq}(\tilde{T}_e) \right] C_{e-ph}(\varepsilon,\varepsilon+\hbar\omega,\omega,\mathbf{p}) \right\} d\omega$$  ,  (17)

$$\frac{1}{\tau_{ph-e}(\omega,\mathbf{p})} = 2\pi \int \frac{D_e(\varepsilon)}{D_{ph}(\omega,\mathbf{p})} \left[ f_\varepsilon^{eq}(\tilde{T}_e) - f_{\varepsilon+\hbar\omega}^{eq}(\tilde{T}_e) \right] C_{e-ph}(\varepsilon,\varepsilon+\hbar\omega,\omega,\mathbf{p}) d\varepsilon .$$  (18)

Eventually, the strongly coupled electron and phonon BTEs (1) and (2) are greatly

simplified into the relaxation time approximation forms:

$$\frac{\partial f_{\mathbf{k}}}{\partial t} + \mathbf{v}_e \cdot \nabla_{\mathbf{r}} f_{\mathbf{k}} = -\frac{f_{\mathbf{k}} - f_{\mathbf{k}}^{eq}(\tilde{T}_e)}{\tau_{e-ph}(\varepsilon)}, \tag{19}$$

$$\frac{\partial n_{\mathbf{Q,p}}}{\partial t} + \mathbf{v}_{ph,\mathbf{p}} \cdot \nabla_{\mathbf{r}} n_{\mathbf{Q,p}} = -\frac{n_{\mathbf{Q,p}} - n_{\mathbf{Q,p}}^{eq}(\tilde{T}_e)}{\tau_{ph-e}(\omega,\mathbf{p})} - \frac{n_{\mathbf{Q,p}} - n_{\mathbf{Q,p}}^{eq}(\tilde{T}_{ph})}{\tau_{\mathrm{U},ph-ph}}, \tag{20}$$

with the relaxation times given by Eq. (12), (17) and (18) separately. The local pseudo-equilibrium distributions for electrons and phonons are the Fermi-Dirac distribution and the Bose-Einstein distribution respectively [49]:

$$f_{\mathbf{k}}^{eq}(\tilde{T}_e) = \frac{1}{\exp\left[(\varepsilon_{\mathbf{k}} - \mu)/k_{\mathrm{B}}\tilde{T}_e\right] + 1} \tag{21}$$

$$n_{\mathbf{Q,p}}^{eq}(\tilde{T}_{ph}) = \frac{1}{\exp\left(\hbar\omega_{\mathbf{Q,p}}/k_{\mathrm{B}}\tilde{T}_{ph}\right) - 1} \tag{22}$$

where $k_{\mathrm{B}}$ is the Boltzmann constant. The chemical potential $\mu$ is very close to Fermi energy $\varepsilon_{\mathrm{F}}$ at the concerned temperature scope nearly lower than one percent of the corresponding Fermi temperature in this work, so that $\mu = \varepsilon_{\mathrm{F}}$ is assumed for simplicity without losing accuracy [49]. The e-ph scattering connects the energy transfer between electrons and phonons so that the electron and phonon BTEs (19) and (20) are not decoupled and the simultaneous solution is essential. If under the pre-assumption that phonons are sufficiently thermalized to be in equilibrium with electrons, the coupled equations are automatically degraded to the electron BTE describing the electron thermal transport.

## 2.2 Computational scheme

In metals, only electrons around the Fermi energy with a width about the thermal energy unit ($k_{\mathrm{B}}T$) will respond to thermal perturbation [42]. Thus, we propose to consider the electron distribution $f_{\mathbf{k}}$ above the Fermi energy and $1-f_{\mathbf{k}}$ below the Fermi energy in the numerical simulation of electron thermal transport, as clearly demonstrated in our previous work [38]. For a compact mathematical expression, we introduce the following distribution function for electron thermal transport:

$$g_{\mathbf{k}} = H\left(\mu - \varepsilon\right) + \left[1 - 2H\left(\mu - \varepsilon\right)\right]f_{\mathbf{k}}, \tag{23}$$

where $H\left(\mu - \varepsilon\right)$ is the Heaviside step function. Thus, the electron BTE for thermal transport is reformulated as

$$\frac{\partial g_{\mathbf{k}}}{\partial t} + \mathbf{v}_e \cdot \nabla_{\mathbf{r}} g_{\mathbf{k}} = -\frac{g_{\mathbf{k}} - g_{\mathbf{k}}^{eq}\left(\tilde{T}_e\right)}{\tau_{e-ph}\left(\varepsilon\right)}, \tag{24}$$

with $g_{\mathbf{k}}^{eq}\left(\tilde{T}_e\right) = \left\{\exp\left(\left|\varepsilon_{\mathbf{k}} - \mu\right|/k_{\mathrm{B}}\tilde{T}_e\right) + 1\right\}^{-1}$.

The local pseudo-temperatures, which are not explicitly involved in the original collision terms, need to be determined in the current form of relaxation time approximation. Especially, whatever the form of scattering terms, the intrinsic conservation laws should be always obeyed during the corresponding scattering channels such as the e-ph and ph-ph scattering. Thus, in terms of the thermal transport, the energy conservation during e-ph scattering and ph-ph scattering process is implemented to determine the electron and phonon pseudo-temperature respectively as:

$$\sum_{\mathbf{Q},\mathrm{p}} \hbar\omega_{\mathbf{Q},\mathrm{p}}\left(-\frac{n_{\mathbf{Q},\mathrm{p}} - n_{\mathbf{Q},\mathrm{p}}^{eq}\left(\tilde{T}_{ph}\right)}{\tau_{\mathrm{U},ph-ph}}\right) = 0, \tag{25}$$

$$\sum_{\mathbf{k}} \left|\varepsilon_{\mathbf{k}} - \mu\right|\left(-\frac{g_{\mathbf{k}} - g_{\mathbf{k}}^{eq}\left(\tilde{T}_e\right)}{\tau_{e-ph}\left(\varepsilon\right)}\right) + \sum_{\mathbf{Q},\mathrm{p}} \hbar\omega_{\mathbf{Q},\mathrm{p}}\left(-\frac{n_{\mathbf{Q},\mathrm{p}} - n_{\mathbf{Q},\mathrm{p}}^{eq}\left(\tilde{T}_e\right)}{\tau_{ph-e}\left(\omega,\mathrm{p}\right)}\right) = 0, \tag{26}$$

and the details are shown in Appendix A. Consequently, the electron and phonon transport equations (24) and (20) combined with Eqs. (25) and (26) are complete for the description of coupled electron and phonon thermal transport process. The calculation of the e-ph coupling constant is thus formulated as

$$G = \frac{\left(\dfrac{\partial E}{\partial t}\right)_{ph-e}}{\tilde{T}_e - \tilde{T}_{ph}} = \frac{\sum_{Q,p} \hbar \omega_{Q,p} \left(-\dfrac{n_{Q,p} - n_{Q,p}^{eq}\left(\tilde{T}_e\right)}{\tau_{ph-e}\left(\omega, p\right)}\right)}{\tilde{T}_e - \tilde{T}_{ph}}, \qquad (27)$$

the contribution to which can be conveniently divided into different phonon branches. Especially, through the description of distribution functions and the construction of e-ph scattering and ph-ph scattering process, the non-equilibrium effects are naturally included.

Furthermore, under the present isotropic assumption, the intensity forms are introduced by multiplying the Eq. (24) and (20) by $v_e |\varepsilon - \mu| D_e(\varepsilon)/4\pi$ and $v_{ph,p} \hbar \omega D_{ph}(\omega, p)/4\pi$ respectively:

$$\frac{\partial I_\varepsilon}{\partial t} + \mathbf{v}_e \cdot \nabla_{\mathbf{r}} I_\varepsilon = -\frac{I_\varepsilon - I_\varepsilon^{eq}\left(\tilde{T}_e\right)}{\tau_{e-ph}\left(\varepsilon\right)}, \qquad (28)$$

$$\frac{\partial \phi_{\omega,p}}{\partial t} + \mathbf{v}_{ph,p} \cdot \nabla_{\mathbf{r}} \phi_{\omega,p} = -\frac{\phi_{\omega,p} - \phi_{\omega,p}^{eq}\left(\tilde{T}_e\right)}{\tau_{ph-e}\left(\omega, p\right)} - \frac{\phi_{\omega,p} - \phi_{\omega,p}^{eq}\left(\tilde{T}_{ph}\right)}{\tau_{U,ph-ph}}. \qquad (29)$$

The physical meaning of electron (phonon) intensity is the flux of energy per unit area, per unit time, per unit solid angle along the direction of electron (phonon) propagation, and per unit energy (frequency) interval around $\varepsilon$ ($\omega$) [50]. The local pseudo-equilibrium intensities are formulated separately as: $I_\varepsilon^{eq}\left(\tilde{T}_e\right) = v_e |\varepsilon - \mu| g_\varepsilon^{eq}\left(\tilde{T}_e\right) D_e(\varepsilon)/4\pi$ and $\phi_{\omega,p}^{eq}\left(\tilde{T}_{ph}\right) = v_{ph,p} \hbar \omega n_{\omega,p}^{eq}\left(\tilde{T}_{ph}\right) D_{ph}(\omega, p)/4\pi$, where the pseudo-temperature is computed by the inverse numerical integration of Eq. (25) and (26) transformed into intensity forms as

$$\sum_p \iint_{4\pi} \left(-\frac{\phi_{\omega,p} - \phi_{\omega,p}^{eq}\left(\tilde{T}_{ph}\right)}{v_{ph,p} \tau_{U-ph-ph}}\right) d\mathbf{\Omega} d\omega = 0, \qquad (30)$$

$$\iint_{4\pi} \left(-\frac{I_\varepsilon - I_\varepsilon^{eq}\left(\tilde{T}_e\right)}{v_e \tau_{e-ph}\left(\varepsilon\right)}\right) d\mathbf{\Omega} d\varepsilon + \sum_p \iint_{4\pi} \left(-\frac{\phi_{\omega,p} - \phi_{\omega,p}^{eq}\left(\tilde{T}_e\right)}{v_{ph,p} \tau_{ph-e}\left(\omega, p\right)}\right) d\mathbf{\Omega} d\omega = 0. \qquad (31)$$

The temperature dependence of the spectral relaxation times is considered in the numerical solution.

To sum up, we obtain the intensity forms of the coupled electron and phonon BTEs for numerical solution. Once the electron and phonon intensities are resolved, the e-ph coupling constant $G$, the local electron and phonon energy density $E_e(t,\mathbf{r})$, $E_{ph}(t,\mathbf{r})$ with the respective local temperature $T_e$ and $T_{ph}$ are thus calculated:

$$G = \frac{\left(\dfrac{\partial E}{\partial t}\right)_{ph-e}}{\tilde{T}_e - \tilde{T}_{ph}} = \frac{\sum_{\mathrm{p}} \iint_{4\pi} \left( -\dfrac{\phi_{\omega,\mathrm{p}} - \phi_{\omega,\mathrm{p}}^{eq}\left(\tilde{T}_e\right)}{v_{ph,\mathrm{p}} \tau_{ph-e}\left(\omega,\mathrm{p}\right)} \right) d\mathbf{\Omega}\, d\omega}{\tilde{T}_e - \tilde{T}_{ph}} , \tag{32}$$

$$E_e\left(t,\mathbf{r}\right) = \iint_{4\pi} \frac{I_\varepsilon}{v_e} d\mathbf{\Omega}\, d\varepsilon = \iint_{4\pi} \frac{I_\varepsilon^{eq}\left(T_e\right)}{v_e} d\mathbf{\Omega}\, d\varepsilon , \tag{33}$$

$$E_{ph}\left(t,\mathbf{r}\right) = \sum_{\mathrm{p}} \iint_{4\pi} \frac{\phi_{\omega,\mathrm{p}}}{v_{ph,\mathrm{p}}} d\mathbf{\Omega}\, d\omega = \sum_{\mathrm{p}} \iint_{4\pi} \frac{\phi_{\omega,\mathrm{p}}^{eq}\left(T_{ph}\right)}{v_{ph,\mathrm{p}}} d\mathbf{\Omega}\, d\omega . \tag{34}$$

DOM scheme is adopted for the numerical solution, the development and validity of which have been demonstrated in our previous work [40, 41]. Additionally, the Gauss-Legendre (G-L) quadrature is adopted for the numerical integration over the electron energy, phonon frequency and angular variable due to its high efficiency as shown in Appendix B.

### 2.3 Model verification

The coupled electron and phonon BTEs in Section 2.1 and 2.2 will be verified by numerical modeling of the ultrafast dynamics process in femtosecond pump-probe experiments. A direct comparison with the experiment is difficult because the probed signal such as the reflectance change of metal surface includes the contribution from both electrons and phonons with the ratio to each other dependent on the probe laser and material. As the validation of the relaxation time approximation of the e-ph coupling in the present model is the key point, we compare the present coupling model to the full integral treatment of scattering terms in Ono's model [34]. The widely used continuum models like two-temperature model (TTM) and four-temperature model (FTM) are also included for comparison.

Neglecting the drift term is often adopted for simplicity when investigating the ultra-fast dynamics in femtosecond pump-probe experiments since the characteristic time scale of coupling and energy transfer between electrons and phonons is very short [34]. The initial condition, electron Fermi window, phonon dispersion relation and the expression of e-ph coupling function in Eq. (15) for aluminum are all chosen from the reference [34]. The Grüneisen parameters in three-phonon Umklapp scattering term for different phonon polarizations are adopted separately as: $\gamma_{TA}$=2.21, $\gamma_{LA}$=2.21 [31]. In DOM scheme for numerical solution, the angular and spatial variables vanish in this case due to the negligible treatment of the drift term. The discretization of the electron energy and phonon spectrum is based on the abscissae of the G-L quadrature that are applied with 96 and 80 points for electron and phonon respectively. The time step is set to be 1 fs.

The excess electron and phonon energy density defined as $EX_e = E_e(t) - E_e(T_0)$, $EX_{ph} = E_{ph}(t) - E_{ph}(T_0)$ separately with the reference temperature $T_0 = 290$ K are shown in the Figure 1. The electron energy density decreases with time accompanied by the increase of phonon energy density due to the energy transfer between each other.

Ono's work demonstrates that the energy relaxation in the continuum models with the quasi-equilibrium treatment such as TTM and FTM is obviously faster than that in his BTE model [34]. Although the present result is slightly faster than that of Ono's model due to the assumption of weak deviation from the equilibrium state, the agreement is generally very good in contrast to the TTM and FTM. For some femtosecond pump-probe experiments where electrons are highly excited to be thousands of Kelvin instantly, the higher-order non-equilibrium effects might be relevant and the integral treatment of full scattering term is more preferable. Nevertheless, in most transport cases within the linear response regime, the present treatment of e-ph coupling by relaxation time approximation provides a more practical avenue. It represents a more accurate theoretical description comparing to continuum model whereas simpler one comparing to BTE with full scattering term. In summary, the comparison of the results by our relaxation-time-approximation model with that by the Ono's model provides a solid validation of our e-ph coupling model.

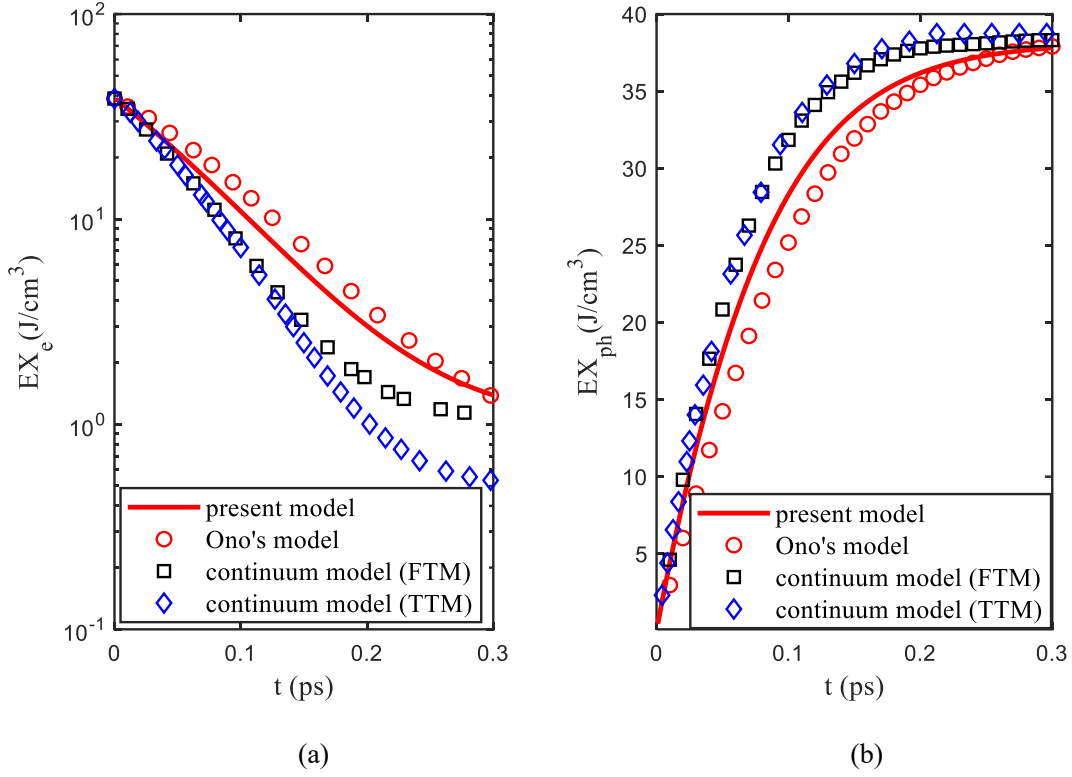

(a)                                         (b)

Figure 1. Time-dependent excess electron and phonon energy density: (a) excess electron energy density; (b) excess phonon energy density; The solid line denotes the result of the present coupling model whereas hollow circles, squares and diamonds are that of Ono's model, FTM and TTM respectively [34].

## 3. Results and discussion

In this section, we will apply the validated e-ph coupling model to study the e-ph coupling constant for four kinds of metals including aluminum, copper, silver and gold. The influences on the e-ph coupling constant by the temporal non-equilibrium between different phonon branches and the spatial non-equilibrium of electrons are investigated in sub-section 3.1 and 3.2, respectively.

The electron band structure is approximated as the free electron model and electron Fermi window is adopted as [$\varepsilon_F$-15$k_B T$, $\varepsilon_F$+15$k_B T$] with $T$ = 500 K. The phonon dispersion relation along [0 0 1] direction is used to represent the dispersion in the first

Brillouin zone, and is fitted by an empirical power law expression: $\omega(q) = B_{4,\text{p}}q^4 + B_{3,\text{p}}q^3 + B_{2,\text{p}}q^2 + B_{1,\text{p}}q$ with fitting parameters $B_{1,\text{p}}$, $B_{2,\text{p}}$, $B_{3,\text{p}}$ and $B_{4,\text{p}}$. This expression for different metals is referred from experimental data [51-54] or theoretic calculation [55] as summarized in is dominated by LA phonons [10].

Table 1. The non-dimensional parameter is expressed as $q = Q/Q_{\max}$ with $Q$ the phonon wave vector along [0 0 1] direction and $Q_{\max} = 2\pi/a$, $a$ being the cubic lattice constant. Transverse acoustic (TA) polarization and longitudinal acoustic (LA) polarization are included for calculation. The Grüneisen parameters for different phonon branches are chosen to be the same [31, 33, 56] as summarized in

Table 2. Besides, the e-ph coupling function is related to the Eliashberg function and approximated as $C_{e-ph}(\varepsilon, \varepsilon', \omega, \text{p}) \simeq \sqrt{\varepsilon_{\text{F}}/\varepsilon}\, \alpha^2 F(\omega, \text{p})$ for the convenience of considering the deviation of electron energy from the Fermi level [34]. Furthermore, an empirical expression originally derived in the low-frequency limit is adopted for the Eliashberg function in the whole frequency spectrum: $\alpha^2 F(\omega, \text{p}) = \lambda_{\text{p}} n (\omega/\omega_{\max,\text{p}})^2/2$ with $n$ equal to 2 for a clean bulk crystal [57]. The mass enhancement parameter $\lambda_{\text{p}}$ for different polarizations can be inversely determined as $\lambda_{\text{p}} = 2\int_0^{\omega_{\max,\text{p}}} \frac{\alpha^2 F(\omega, \text{p})}{\omega} d\omega$ once the Eliashberg function is obtained [58]. As the overall mass enhancement parameter $\lambda = 2\lambda_{\text{TA}} + \lambda_{\text{LA}}$ is usually given by *ab initio* theoretical calculation [31-33] or experimental measurement [58], the contribution from different phonon branches is often not given. Thus the ratio of different branches is referred from the empirical calculation for aluminum [34] and is approximated to be $\lambda_{\text{LA}}$:$\lambda_{\text{TA}} \approx 2$:1 as given in

Table 2. This value is consistent with the theoretical analysis that the e-ph coupling

Table 1. Phonon dispersion relation along [0 0 1] direction for different metals including aluminum, silver, copper and gold at 300 K

| | TA branch(rad/s) | LA branch(rad/s) |
|---|---|---|
| Al | $6.750\times10^{12}q^4 - 3.204\times10^{13}q^3$ $+9.578\times10^{12}q^2 + 5.147\times10^{13}q$ | $6.208\times10^{13}q^4 - 1.216\times10^{14}q^3$ $+2.226\times10^{13}q^2 + 9.825\times10^{13}q$ |
| Ag | $8.1289\times10^{12}q^4 - 2.4607\times10^{13}q^3$ $+4.4373\times10^{12}q^2 + 3.3748\times10^{13}q$ | $1.5349\times10^{13}q^4 - 3.8537\times10^{13}q^3$ $+0.7202\times10^{12}q^2 + 5.4066\times10^{13}q$ |
| Cu | $1.0452\times10^{13}q^4 - 3.5385\times10^{13}q^3$ $+7.6611\times10^{12}q^2 + 4.9674\times10^{13}q$ | $4.0666\times10^{12}q^4 - 2.643\times10^{13}q^3$ $-6.351\times10^{12}q^2 + 7.4353\times10^{13}q$ |
| Au | $1.1451\times10^{12}q^4 - 1.3922\times10^{13}q^3$ $+7.7833\times10^{12}q^2 + 2.2398\times10^{13}q$ | $4.0651\times10^{12}q^4 - 8.6902\times10^{12}q^3$ $-2.1783\times10^{13}q^2 + 5.5651\times10^{13}q$ |

Table 2. Summary of electron Fermi energy $\varepsilon_F$, cubic lattice constant $a$, overall mass enhancement parameter $\lambda$ including the contribution from TA polarization $\lambda_{TA}$ and LA porization $\lambda_{LA}$, and Grüneisen parameters $\gamma_{TA}$, $\gamma_{TA}$ for TA, LA branch for different metals at 300 K

| | $\varepsilon_F$/eV [49] | $a$/Å [49] | $\lambda$ | $\lambda_{TA}$ | $\lambda_{LA}$ | $\gamma_{TA}$ | $\gamma_{LA}$ |
|---|---|---|---|---|---|---|---|
| Al | 11.63 | 4.05 | 0.45 | 0.12 | 0.21 | 2.21 | 2.21 |
| Ag | 5.48 | 4.09 | 0.12 | 0.03 | 0.06 | 2.31 | 2.31 |
| Cu | 7.00 | 3.61 | 0.18 | 0.045 | 0.09 | 1.94 | 1.94 |
| Au | 5.51 | 4.08 | 0.15 | 0.04 | 0.07 | 2.62 | 2.62 |

In order to validate the parameters adopted in our model, the electron thermal conductivity $\kappa_e = \int_{Fermi\ window} \left(\varepsilon - \varepsilon_F\right) \dfrac{df^{eq}}{dT} v_e^2 \tau_{e-ph}\left(\varepsilon\right) D_e\left(\varepsilon\right) d\varepsilon \Big/ 3$, phonon thermal conductivity $\kappa_{ph} = \sum_p \int_0^{\omega_{max,p}} \hbar\omega \dfrac{dn^{eq}}{dT} v_{ph,p}^2 \left(\tau_{ph-e}^{-1} + \tau_{ph-ph}^{-1}\right)^{-1} D_{ph}\left(\omega,p\right) d\omega \Big/ 3$ based on the    kinetic    theory    [42]    and    the    e-ph    coupling    constant

$$G = 3\hbar \int_{\substack{Fermi \\ window}} (\varepsilon - \varepsilon_{\mathrm{F}}) \frac{df^{eq}}{dT} D_e(\varepsilon) d\varepsilon * 2 \int_0^{\omega_{\max,\mathrm{p}}} \frac{\alpha^2 F(\omega,\mathrm{p})}{\omega} \omega^2 d\omega \Big/ \pi k_{\mathrm{B}} T$$    based    on    the

Allen's formula [17] are separately calculated for aluminum, silver, copper and gold, as summarized in

Table 3. The overall thermal conductivity summing the contribution from electron and phonon as $\kappa = \kappa_e + \kappa_{\mathrm{ph}}$ is also shown. We compare the results of the present calculation with those by *ab initio* calculation [31-33] and experimental measurements [49, 58, 59]. For the overall thermal conductivity of Al, there exists an appreciable difference between the experimental data and the present result, which may arise from the calculation of the Fermi velocity [60]. For the phonon thermal conductivity of Al and Cu, the results by the present work are a little lower than that by *ab initio* method. It might be attributed to the power law approximation of Eliashberg function in the whole spectrum, which may slightly overpredict the ph-e scattering rate for the medium-high frequency. Thus, the corresponding phonon thermal conductivity of Al and Cu with higher Debye temperature is underestimated slightly. Generally, the e-ph coupling constant and overall thermal conductivity calculated by the present work agree with the *ab initio* results and experimental data, which validates the input parameters adopted in our model.

Table 3. Summary of electron thermal conductivity $\kappa_e$, phonon thermal conductivity $\kappa_{\mathrm{ph}}$, overall thermal conductivity $\kappa$ and e-ph coupling constant $G$ calculated by the present work in comparison to *ab initio* results [31-33] or experimental data [49, 58, 59] for different metals at 300 K

|  |  | $\kappa_e$(W/m/K) | $\kappa_{\mathrm{p}}$(W/m/K) | $\kappa$(W/m/K) | $G$(W/m³/K) |
|---|---|---|---|---|---|
|  | present calculation | 335.24 | 3.91 | 339.15 | $3.62 \times 10^{17}$ |
| Al | *ab initio* | 232.53 [33] | 8.95 [33] | 241.49 [33] | $5.38\times10^{17}$ [31] |
| | | 246 [31] | 5.8 [32], 6 [31] | 252 [31] | |
| | experiment | - | - | 237 [49] | $2.45\times10^{17}$ [59] |
| Ag | present calculation | 404.94 | 4.94 | 409.88 | $2.01\times10^{16}$ |
| | *ab initio* | 450.86 [33] | 5.69 [33] | 456.55 [33] | $3.0\times10^{16}$ [31] |
| | | 370 [31] | 5.2 [32], 4 [31] | 374 [31] | |
| | experiment | - | - | 429 [49] | |
| Cu | present calculation | 388.55 | 10.26 | 398.81 | $7.25\times10^{16}$ |
| | *ab initio* | 361.32 [33] | 17.42 [33] | 378.74 [33] | |
| | | | 16.9 [32] | | |
| | experiment | - | - | 401 [49] | $\sim1\times10^{17}$ [58] |
| Au | present calculation | 326.88 | 4.05 | 330.93 | $1.92\times10^{16}$ |
| | *ab initio* | 273.45 [33] | 2.80 [33] | 276.25 [33] | $2.2\times10^{16}$ [31] |
| | | 276 [31] | 2.6 [32], 2 [31] | 278 [31] | |
| | experiment | - | - | 317 [49] | $2.9\times10^{16}$ [59] |

## 3.1  Temporal non-equilibrium effect on e-ph coupling constant

In this sub-section, the e-ph coupling constant in ultra-fast dynamics is calculated. For simplicity, the drift term is neglected as a first step. The initial condition is set that electrons are assumed in equilibrium at 980 K with phonons undisturbed at room temperature (300 K). This choice of initial condition is to approach the condition in femtosecond pump-probe experiments as a first step. The time step for aluminum, silver, copper and gold is adopted to be 2 fs, 10 fs, 2 fs and 8 fs separately. The number of abscissae of the G-L quadrature is chosen as 96 and 80 points for electron energy and phonon spectrum respectively. Thus, the time-dependent e-ph coupling constant for different metals is computed based on Eq. (32) as shown in Figure 2, where the electron and phonon temperature are also displayed. The energy transfer occurs between electrons and phonons, which is clearly shown by the decrease of electron temperature and increase of phonon temperature until the temperature difference between each other vanishes. The e-ph coupling constant predicted by Allen's theory is constant within the temperature scope concerned in the present work where the excitation of d band electrons is not necessarily considered. In comparison, the e-ph coupling constant calculated by the present model remains nearly constant little lower than that predicted by Allen's theory at the initial stage, rapidly decreasing at some moment and finally reduces to the value one order of magnitude smaller than that of Allen's theory. The temperature difference between electron and phonon is still appreciable (27.2 K, 5.6 K, 10.8 K and 7.1 K for aluminum, silver, copper and gold separately) when the e-ph coupling constant reduces to half of Allen's theoretical value. Besides, the occurrence instant of this rapid reduction is different for different metals.

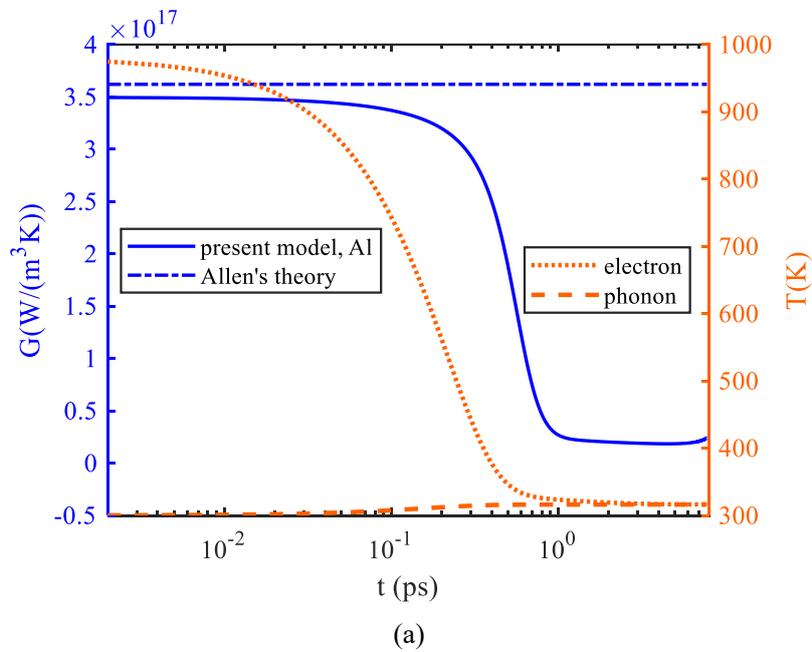

(a)

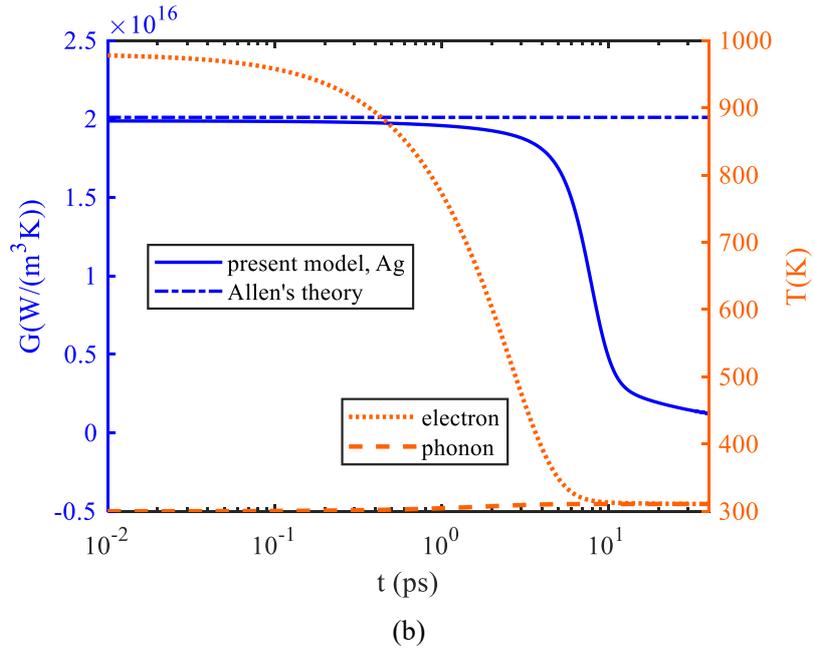

(b)

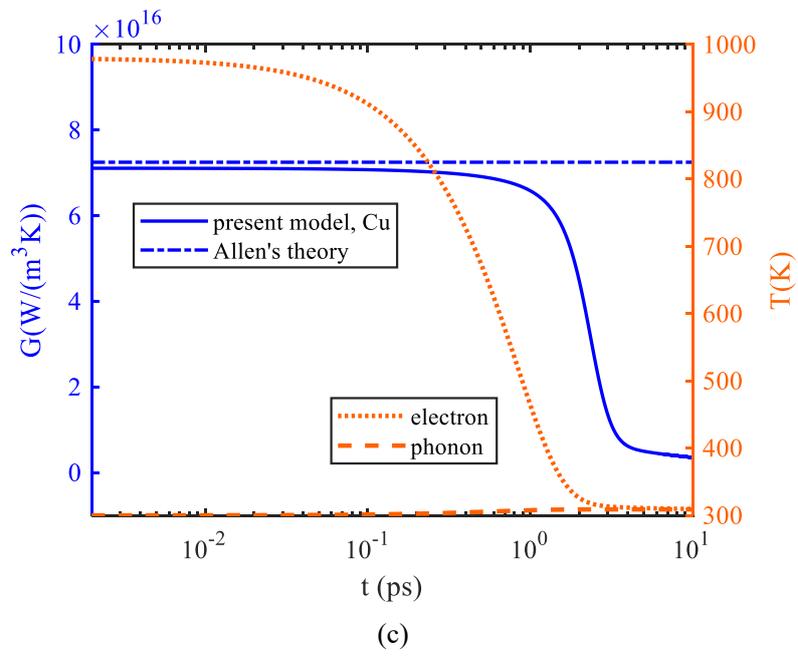

(c)

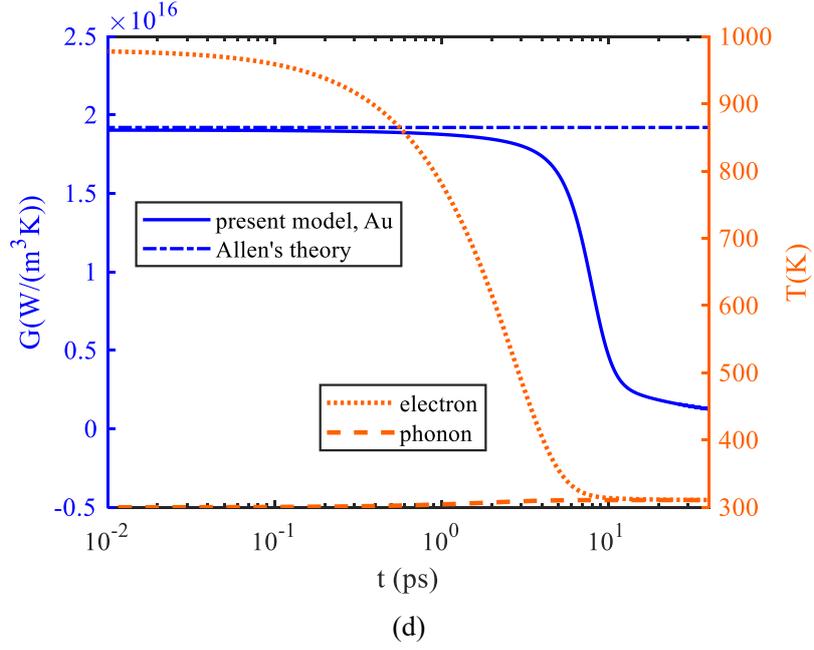

(d)

Figure 2. Time-dependent e-ph coupling constant and electron, phonon temperature in ultrafast dynamics: (a) for aluminum; (b) for silver; (c) for copper; (d) for gold. The solid blue line represents the e-ph coupling constant calculated by the present model whereas the blue dash dot line denotes that from Allen's theory [17]; The electron and phonon temperature are marked by the orange dot line and dash line separately.

For the further understanding of this reduction, the contribution from different phonon branches to the e-ph coupling constant is displayed in Figure 3 for Ag. This reduction is almost attributed to LA branch whereas the e-ph coupling constant due to TA branch remains nearly constant. The present trend is also applicable for other metals including aluminum, copper and gold, and is not shown here to avoid repetition. The e-ph coupling constant contributed by LA branch even reduces to be negative, which means that LA phonons transfer energy to electrons. Such anomalous phenomena of energy back-flow has been also demonstrated in the Ono's model [34].

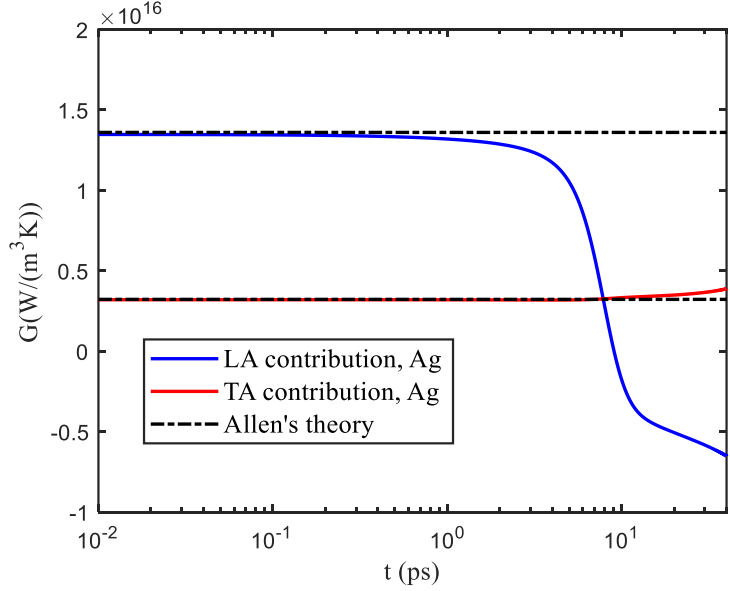

Figure 3. The contribution from TA and LA branch to the e-ph coupling constant for silver: the blue and red line represent the result of present model contributed by LA and TA branch respectively whereas the dash dot line denotes that of Allen's theory [34].

Furthermore, in order to investigate how the energy back-flow occurs, the TA and LA phonon energy density change rates by the e-ph scattering and ph-ph scattering are defined as $\left(\partial E_{ph,\text{TA}}/\partial t\right)_{e-ph}$, $\left(\partial E_{ph,\text{LA}}/\partial t\right)_{e-ph}$, $\left(\partial E_{ph,\text{TA}}/\partial t\right)_{ph-ph}$ $\left(\partial E_{ph,\text{LA}}/\partial t\right)_{ph-ph}$ separately. Through integrating the corresponding scattering term, the energy density change rates are computed and shown in Figure 4(a). Figure 4(b) shows the excess energy density of electron and phonon at different polarizations with the reference temperature of 300 K. During the initial stage to 0.3 ps, the energy exchange by the e-ph scattering dominates and the strength of electron-LA phonon (e-LA) scattering is stronger than that of electron-TA phonon (e-TA) scattering. It is clearly shown in Figure 4(b) that the energy increase of LA phonon is much faster than that of TA phonon such that non-equilibrium occurs between different phonon branches. This non-equilibrium effect continually enlarges from 0.3 ps to 2 ps due to the continuous dominance of e-ph scattering though ph-ph scattering gradually works pushing LA and TA phonon to be equilibrium. Afterwards, during the next 7 ps, ph-ph scattering gradually dominates and thus the non-equilibrium between TA and LA phonons is weakened but still exists. Considering the continually decreasing amount of energy exchange by e-ph scattering, the energy transfer from electrons to LA phonons rapidly reduces to be negative around 9 ps. In other words, energy back-flow takes place, the clear signature of which is revealed in the inset of Figure 4(a).

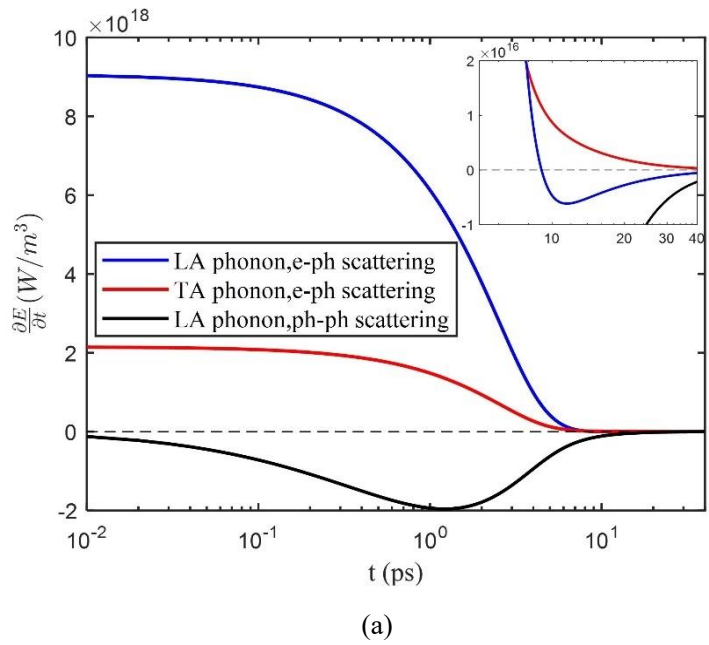

(a)

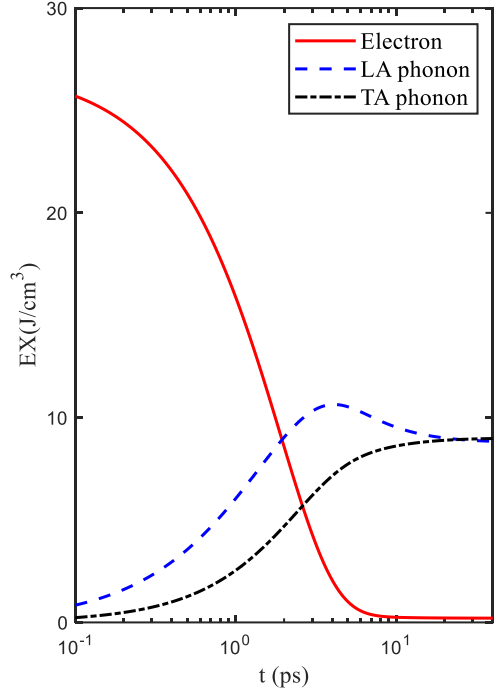

(b)

Figure 4. Time-dependent phonon energy density change rate and the excess energy density of electron and phonon for silver: (a) phonon energy density change rate; the blue and black solid line represent the result of LA phonon by e-ph scattering and ph-ph scattering separately whereas the red line denotes that of TA phonon by e-ph scattering, and the dash line is shown for eye guidance; The energy density change rate of TA phonon by ph-ph scattering is correspondingly inverse with that of LA phonon due to the energy conservation principle during ph-ph scattering process; The inset is the magnified figure of period from 5 ps to 40 ps; (b) the excess energy density of electron and phonon; The red line represents the result of electron whereas the blue dash and black dash dot line denote that of LA phonon and TA phonon respectively.

In general, the intrinsically different dispersion relation for TA and LA phonons and the different coupling strength for e-LA and e-TA scattering represented by the coupling function in Eq. (15) will induce non-equilibrium between different phonon polarizations. This will further make the ph-ph scattering gradually work pushing the equilibrium between TA and LA phonons. Considering these effects, the temporal non-equilibrium between different phonon branches always exists. Finally, energy back-flow occurs and the e-ph coupling constant rapidly decreases. The present exploration of the time-dependent e-ph coupling constant by our model is significant for the description of electron and phonon coupling process in the femtosecond pump-probe experiments. The use of an invariable e-ph coupling constant in the continuum TTM model requires further reexamination in the near future.

Besides, the different types of initial condition are investigated. First of all, the Gaussian-type initial distribution for electron [34] can be straightforwardly incorporated, the influence of which on the e-ph coupling constant comparing to the currently pre-assumed high temperature condition is checked and shown in the Figure 5(a). The Gaussian-type initial condition that approximately amounts to an effective electron temperature of 980 K has no effect on the result of e-ph coupling constant. Moreover, the influence by different values of pre-assumed electron temperature is also investigated in Figure 5(b). The e-ph coupling constant is nearly independent on electron temperature lower than 1000 K, which agrees with the conclusion in the *ab initio* calculation [30]. It clearly shows that the occurrence of the reduction of e-ph coupling constant is minorly influenced by the choice of different electron initial temperatures. This may arise from the temperature dependence of relaxation time. These results further validate the present model with more general application.

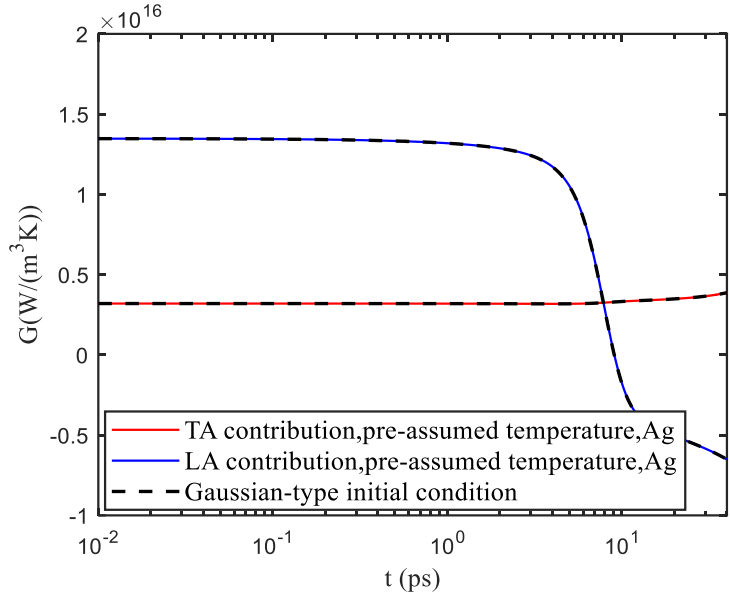

(a)

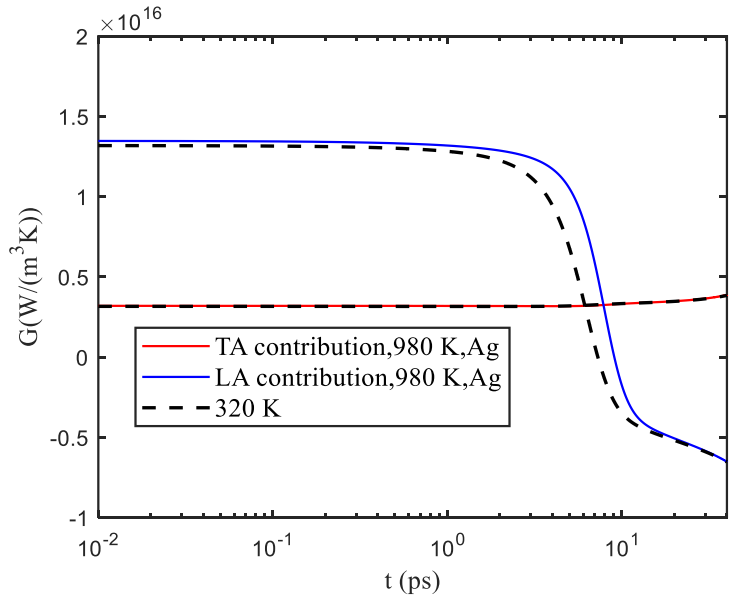

(b)

Figure 5. The influence of different types of electron initial condition on the e-ph coupling constant for silver: (a) Gaussian-type initial state vs. the pre-assumed high temperature condition; the blue and red line represent the results of the pre-assumed temperature condition whereas the black dash line denotes that of Gaussian-type initial state; (b) the different values of pre-assumed initial electron temperature; the blue and red line represent the results of the initial electron temperature of 980 K whereas the black dash line denotes that of initial electron temperature of 320 K.

*3.2 Spatial non-equilibrium effect on e-ph coupling constant*

In this sub-section, the e-ph coupling constant is calculated when the characteristic length of metals is reduced to nanoscale size. The quantum effect, which might be significant below several nanometers, will not be considered in the present work as a first step. The cross-plane electron and phonon coupling transport through a thin gold film is simulated with the isothermal boundary and periodic boundary condition exerted on $x$ direction and $y$ direction respectively. $z$ direction is omitted for simplicity. The temperature of $T_h$=310 K and $T_c$=290 K is assigned on the electron and phonon simultaneously at the left-hand and right-hand sides of $x$ direction respectively. The thicknesses of 400 nm, 80 nm and 5 nm are respectively simulated with a width of 2 nm for the lateral periodic boundary. In numerical solution, the number of abscissae of the G-L quadrature is chosen as 48 and 48 points for electron energy and phonon spectrum respectively and 32 points for both the discretization of azimuth and polar angle. The number of spatial grids is 201, 81 and 41 for thickness of 400 nm, 80 nm and 5 nm separately with 3 nodes for $y$ direction. Therefore, the e-ph coupling constant and the local non-dimensional temperature of electron and phonon are calculated and shown in Figure 6. When the temperatures of electron and phonon gradually coincide as clearly shown in the middle region in Figure 6(a) and (b), the amount of energy exchange between electrons and phonons tends to be zero. The use of expression Eq. (32) to calculate e-ph coupling constant is meaningless and thus the result of the e-ph coupling constant is not displayed in this middle region.

The mean free path (MFP) of phonon is about one order of magnitude smaller than that of electron in metals [20]. Thus, electrons in the thin film are more difficult to be sufficiently thermalized with the isothermal boundaries. It is manifested by a larger temperature jump of electron than that of phonon at the boundary, which further increases as the film thickness decreases. Consequently, non-equilibrium between electrons and phonons exists, and then an energy exchange occurs. However, the energy exchange (shown in the e-ph coupling constant) is weakened compared to the Allen's theoretical calculation, which shall be attributed to the spatial non-equilibrium effect of electrons. In other words, electrons are more likely to reach the boundary before they exchange energy with phonons so that the e-ph coupling constant decreases near the boundary. In addition, the e-ph coupling constant further decreases by the increase of this spatial non-equilibrium effect as the film thickness decreases. The other metals including aluminum, silver and copper exhibit the similar results not shown here. This exploration will promote the fundamental understanding of electron and phonon coupling at very small scale. Generally, the non-equilibrium effects of electrons and phonons are increasingly important at nanoscale. As a result, the electron drag and phonon drag might play a non-negligible role at this small scale besides at low temperature. It is rarely touched in the literature and requires more studies in near future.

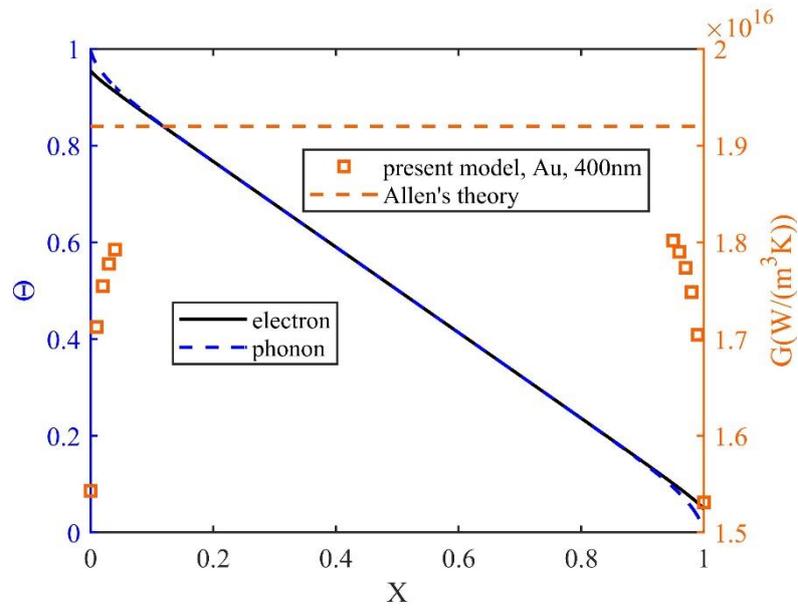

(a)

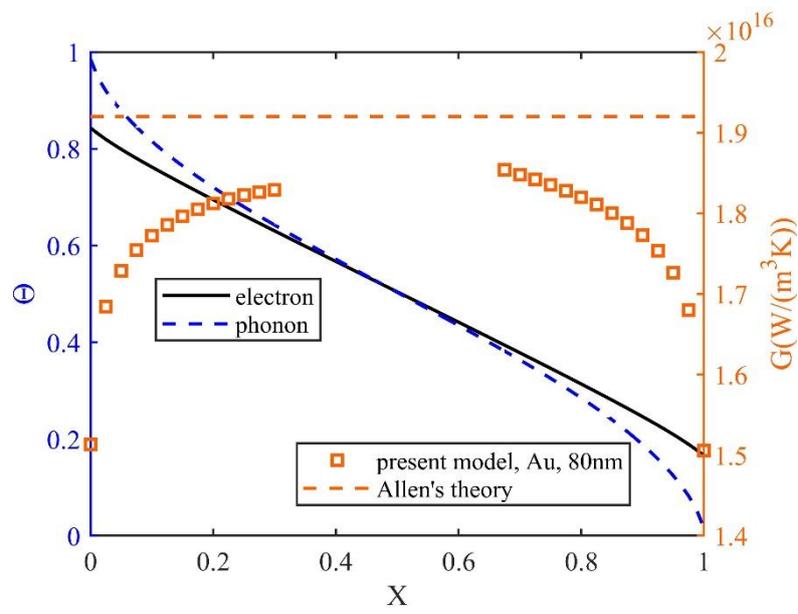

(b)

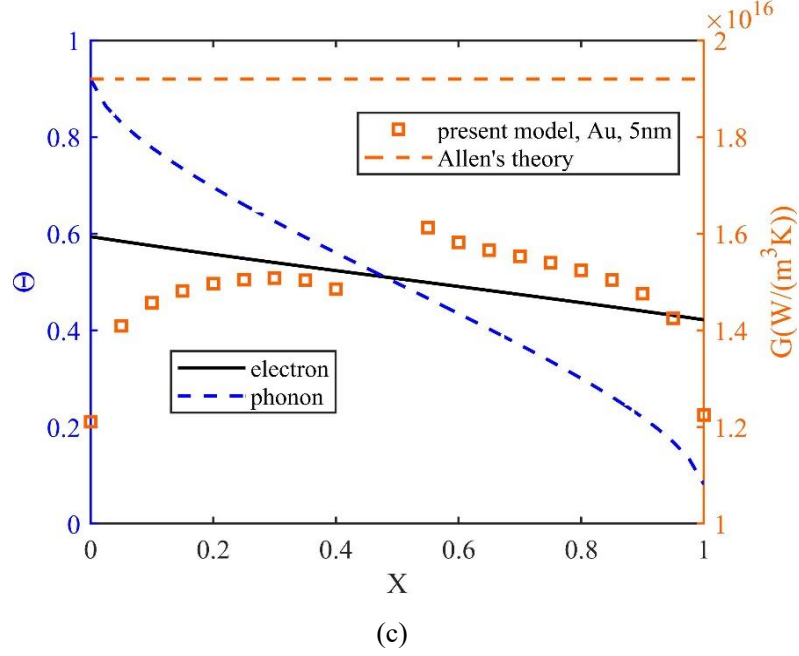

(c)

Figure 6. The e-ph coupling constant and the non-dimensional temperature $\Theta = (T - T_c)/(T_h - T_c)$ of electron and phonon for the thin gold film of different thickness: (a) 400 nm; (b) 80 nm; (c) 5 nm. The hollow squares represent the e-ph coupling constant calculated by the present model whereas the orange dash line denotes that from Allen's theory [34]; The non-dimensional electron and phonon temperature are marked by the black solid line and blue dash line separately.

## 4. Conclusions

In the present work, an electron-phonon (e-ph) coupling model is proposed through transforming the original e-ph and ph-e scattering terms in the coupled electron and phonon Boltzmann transport equations into the relaxation time approximation forms. This model is applied to investigate the e-ph coupling constant when the temporal or spatial non-equilibrium effects are important. We demonstrate a verification of the coupling model by numerical modeling of the ultrafast dynamics process in femtosecond pump-probe experiments, which shows generally consistent results with the full integral treatment of scattering terms. The e-ph coupling constant decreases rapidly due to the temporal non-equilibrium effect between different phonon branches. Moreover, the e-ph coupling constant is much reduced by the non-equilibrium of electrons in confined space. As the non-equilibrium effect is intrinsically included, this e-ph coupling model provides a feasible tool for the theoretical description of coupled electron and phonon transport systems at micro- and nanoscale. The present work will promote the fundamental understanding and modeling of electron and phonon coupling process.

**Acknowledgements**

This work is financially supported by NSF of China (No.51621062, U1837602). Our simulations are run on the "Explorer 100" cluster of Tsinghua National Laboratory for Information Science and Technology. The authors appreciate helpful discussions with Y.Y. Guo and G. Yang.

## APPENDIX A: DETERMINATION OF PSEUDO-TEMPERATURES

We consider the zero-order moment equations through multiplying Eqs. (24) and (20) by $|\varepsilon_{\mathbf{k}} - \mu|$ and $\hbar\omega_{\mathbf{Q},p}$ respectively and summing over the wave vector space:

$$\sum_{\mathbf{k}} |\varepsilon_{\mathbf{k}} - \mu| \frac{\partial g_{\mathbf{k}}}{\partial t} + \sum_{\mathbf{k}} \mathbf{v}_e |\varepsilon_{\mathbf{k}} - \mu| \cdot \nabla_{\mathbf{r}} g_{\mathbf{k}} = \sum_{\mathbf{k}} -|\varepsilon_{\mathbf{k}} - \mu| \frac{g_{\mathbf{k}} - g_{\mathbf{k}}^{eq}(\tilde{T}_e)}{\tau_{e-ph}(\varepsilon)}, \qquad (A1)$$

$$\sum_{\mathbf{Q},p} \hbar\omega_{\mathbf{Q},p} \frac{\partial n_{\mathbf{Q},p}}{\partial t} + \sum_{\mathbf{Q},p} \mathbf{v}_{ph,p} \hbar\omega_{\mathbf{Q},p} \cdot \nabla_{\mathbf{r}} n_{\mathbf{Q},p} = \sum_{\mathbf{Q},p} -\hbar\omega_{\mathbf{Q},p} \frac{n_{\mathbf{Q},p} - n_{\mathbf{Q},p}^{eq}(\tilde{T}_e)}{\tau_{ph-e}(\omega,p)} + \sum_{\mathbf{Q},p} -\hbar\omega_{\mathbf{Q},p} \frac{n_{\mathbf{Q},p} - n_{\mathbf{Q},p}^{eq}(\tilde{T}_{ph})}{\tau_{\mathrm{U},ph-ph}}$$

$$. \qquad (A2)$$

Moreover, the energy balance equations require:

$$\frac{\partial E_e}{\partial t} + \nabla \cdot \mathbf{q}_e = -\Psi_{e-ph}, \qquad (A3)$$

$$\frac{\partial E_{ph}}{\partial t} + \nabla \cdot \mathbf{q}_{ph} = \Psi_{e-ph}, \qquad (A4)$$

where $\Psi_{e-ph}$ denotes the energy transfer from electrons to phonons. The heat flux of electrons and phonons is defined as $\mathbf{q}_e = \sum_{\mathbf{k}} \mathbf{v}_e |\varepsilon_{\mathbf{k}} - \mu| g_{\mathbf{k}}$, $\mathbf{q}_{ph} = \sum_{\mathbf{Q},p} \mathbf{v}_{ph,p} \hbar\omega_{\mathbf{Q},p} n_{\mathbf{Q},p}$ separately. Correspondingly, comparing the energy balance equations with the zero-order moment equations, it generally gives to the equations:

$$\sum_{\mathbf{k}} -|\varepsilon_{\mathbf{k}} - \mu| \frac{g_{\mathbf{k}} - g_{\mathbf{k}}^{eq}(\tilde{T}_e)}{\tau_{e-ph}(\varepsilon)} + \sum_{\mathbf{Q},p} -\hbar\omega_{\mathbf{Q},p} \frac{n_{\mathbf{Q},p} - n_{\mathbf{Q},p}^{eq}(\tilde{T}_e)}{\tau_{ph-e}(\omega,p)} = 0 \qquad (A5)$$

$$\sum_{\mathbf{Q},p} -\hbar\omega_{\mathbf{Q},p} \frac{n_{\mathbf{Q},p} - n_{\mathbf{Q},p}^{eq}(\tilde{T}_{ph})}{\tau_{\mathrm{U},ph-ph}} = 0 \qquad (A6)$$

which are actually that (26) and (25) in the main text. In other words, the Eqs. (A5) and (A6) describe the energy conservation principle during the e-ph and ph-ph scattering process, respectively. Thus, these two equations are implemented to calculate the electron and phonon pseudo-temperature.

**APPENDIX B: DOM SCHEME**

A transient two-dimensional scheme is considered whereas the extension to 3D problem is straightforward. For a short abbreviation, $I$ and $\phi$ are used to denote $I_\varepsilon$ and $\phi_{\omega,\mathrm{p}}$ separately. Considering the discretization of the angular and frequency/energy domains by the abscissae of the Gauss-Legendre (G-L) quadrature, the intensity forms of electron and phonon BTEs (28) and (29) reduce to:

$$\frac{\partial I_n^{\theta,\varphi}}{v_{e,n}\partial t} + u_\theta \frac{\partial I_n^{\theta,\varphi}}{\partial x} + \eta_\theta^\varphi \frac{\partial I_n^{\theta,\varphi}}{\partial y} = -\frac{I_n^{\theta,\varphi} - I_n^{eq}\left(\widetilde{T}_e\right)}{\left(v_e \tau_{e-ph}\right)_n} \tag{B1}$$

$$\frac{\partial \phi_{m_\mathrm{p}}^{\theta,\varphi}}{v_{ph,m_\mathrm{p}}\partial t} + u_\theta \frac{\partial \phi_{m_\mathrm{p}}^{\theta,\varphi}}{\partial x} + \eta_\theta^\varphi \frac{\partial \phi_{m_\mathrm{p}}^{\theta,\varphi}}{\partial y} = -\frac{\phi_{m_\mathrm{p}}^{\theta,\varphi} - \phi_{m_\mathrm{p}}^{eq}\left(\widetilde{T}_e\right)}{\left(v_{ph}\tau_{ph-e}\right)_{m_\mathrm{p}}} - \frac{\phi_{m_\mathrm{p}}^{\theta,\varphi} - \phi_{m_\mathrm{p}}^{eq}\left(\widetilde{T}_{ph}\right)}{\left(v_{ph}\tau_{\mathrm{U},ph-ph}\right)_{m_\mathrm{p}}} \tag{B2}$$

where $n = 1, 2, \ldots, N_e$ and $m_\mathrm{p} = 1, 2, \ldots, N_{ph,\mathrm{p}}$ are the discrete electron energy nodes and spectral nodes of phonon polarization p, respectively. $v_{e,n}$ and $v_{ph,m_\mathrm{p}}$ are the module of electron drift velocity and phonon group velocity of different branches. $\theta = 1, 2, \ldots, N_\theta$ and $\varphi = 1, 2, \ldots, N_\varphi$ are the discrete nodes for the polar angle $[0, \pi]$ and semi-azimuth angle $[0, \pi]$ separately with $\eta_\theta^\varphi = \sqrt{1-u_\theta^2}\cos\left(\left(1+u_\varphi\right)\pi/2\right)$, $u_\theta$ and $u_\varphi$ being the corresponding abscissae of G-L quadrature.

Subsequently, the implicit and first-order upwind scheme is applied to the temporal and spatial discretization. Considering the sign of $u_\theta$ and $\eta_\theta^\varphi$, the Eqs. (B1) and (B2) are further written into the discrete forms by the general format:

$$\frac{I_{n,i,j}^{\theta,\varphi,t+1} - I_{n,i,j}^{\theta,\varphi,t}}{v_{e,n}\Delta t} + \frac{u_\theta + |u_\theta|}{2}\frac{I_{n,i,j}^{\theta,\varphi,t+1} - I_{n,i-1,j}^{\theta,\varphi,t+1}}{\Delta x} + \frac{u_\theta - |u_\theta|}{2}\frac{I_{n,i+1,j}^{\theta,\varphi,t+1} - I_{n,i,j}^{\theta,\varphi,t+1}}{\Delta x}$$
$$+ \frac{\eta_\theta^\varphi + |\eta_\theta^\varphi|}{2}\frac{I_{n,i,j}^{\theta,\varphi,t+1} - I_{n,i,j-1}^{\theta,\varphi,t+1}}{\Delta y} + \frac{\eta_\theta^\varphi - |\eta_\theta^\varphi|}{2}\frac{I_{n,i,j+1}^{\theta,\varphi,t+1} - I_{n,i,j}^{\theta,\varphi,t+1}}{\Delta y} = -\frac{I_{n,i,j}^{\theta,\varphi,t+1} - I_{n,i,j}^{eq,t+1}\left(\widetilde{T}_e\right)}{\left(v_e\tau_{e-ph}\right)_n},$$

$$\tag{B3}$$

$$
\frac{\phi_{m_p,i,j}^{\theta,\varphi,t+1} - \phi_{m_p,i,j}^{\theta,\varphi,t}}{v_{ph,m_p}\Delta t} + \frac{u_\theta + |u_\theta|}{2}\frac{\phi_{m_p,i,j}^{\theta,\varphi,t+1} - \phi_{m_p,i-1,j}^{\theta,\varphi,t+1}}{\Delta x} + \frac{u_\theta - |u_\theta|}{2}\frac{\phi_{m_p,i+1,j}^{\theta,\varphi,t+1} - \phi_{m_p,i,j}^{\theta,\varphi,t+1}}{\Delta x}
$$

$$
+ \frac{\eta_\theta^\varphi + |\eta_\theta^\varphi|}{2}\frac{\phi_{m_p,i,j}^{\theta,\varphi,t+1} - \phi_{m_p,i,j-1}^{\theta,\varphi,t+1}}{\Delta y} + \frac{\eta_\theta^\varphi - |\eta_\theta^\varphi|}{2}\frac{\phi_{m_p,i,j+1}^{\theta,\varphi,t+1} - \phi_{m_p,i,j}^{\theta,\varphi,t+1}}{\Delta y} \qquad . \qquad \text{(B4)}
$$

$$
= -\frac{\phi_{m_p,i,j}^{\theta,\varphi,t+1} - \phi_{m_p,i,j}^{eq,t+1}\left(\tilde{T}_e\right)}{\left(v_{ph}\tau_{ph-e}\right)_{m_p}} - \frac{\phi_{m_p,i,j}^{\theta,\varphi,t+1} - \phi_{m_p,i,j}^{eq,t+1}\left(\tilde{T}_{ph}\right)}{\left(v_{ph}\tau_{\mathrm{U},ph-ph}\right)_{m_p}}
$$

In Eqs. (B3) and (B4), $i = 1, 2, \ldots, N_x$ and $j = 1, 2, \ldots, N_y$ denote the spatial notes in the $x$ direction and $y$ direction, respectively whereas $t = 0, 1, 2, \ldots, N_t$ represents the temporal nodes. Thus, the discrete electron and phonon intensities are derived from Eqs. (B3) and (B4) separately:

$$
I_{n,i,j}^{\theta,\varphi,t+1} = \frac{I_{n,i,j}^{\theta,\varphi,t} + \alpha_{e-ph}I_{n,i,j}^{eq,t+1}\left(\tilde{T}_e\right) + \frac{c_n^u + |c_n^u|}{2}I_{n,i-1,j}^{\theta,\varphi,t+1} + \frac{|c_n^u| - c_n^u}{2}I_{n,i+1,j}^{\theta,\varphi,t+1} + \frac{c_n^\eta + |c_n^\eta|}{2}I_{n,i,j-1}^{\theta,\varphi,t+1} + \frac{|c_n^\eta| - c_n^\eta}{2}I_{n,i,j+1}^{\theta,\varphi,t+1}}{1 + \alpha_{e-ph} + |c_u| + |c_\eta|}
$$

$$
, \qquad\qquad\qquad \text{(B5)}
$$

$$
\phi_{m_p,i,j}^{\theta,\varphi,t+1} = \frac{\phi_{m_p,i,j}^{\theta,\varphi,t} + \Phi_{m_p,i,j}^{eq,t+1} + \frac{c_{m_p}^u + |c_{m_p}^u|}{2}\phi_{m_p,i-1,j}^{\theta,\varphi,t+1} + \frac{|c_{m_p}^u| - c_{m_p}^u}{2}\phi_{m_p,i+1,j}^{\theta,\varphi,t+1} + \frac{c_{m_p}^\eta + |c_{m_p}^\eta|}{2}\phi_{m_p,i,j-1}^{\theta,\varphi,t+1} + \frac{|c_{m_p}^\eta| - c_{m_p}^\eta}{2}\phi_{m_p,i,j+1}^{\theta,\varphi,t+1}}{1 + \alpha_{ph-e} + \alpha_{ph-ph} + |c_{m_p}^u| + |c_{m_p}^\eta|}
$$

$$
, \qquad\qquad\qquad \text{(B6)}
$$

with:

$$
\begin{aligned}
c_n^u &= v_{e,n}u_\theta\Delta t \big/ \Delta x \\
c_n^\eta &= v_{e,n}\eta_\theta^\varphi\Delta t \big/ \Delta y \\
c_{m_p}^u &= v_{ph,m_p}u_\theta\Delta t \big/ \Delta x \\
c_{m_p}^\eta &= v_{ph,m_p}\eta_\theta^\varphi\Delta t \big/ \Delta y \\
\alpha_{e-ph} &= \Delta t \big/ \tau_{e-ph,n} \\
\alpha_{ph-e} &= \Delta t \big/ \tau_{ph-e,m_p} \\
\alpha_{ph-ph} &= \Delta t \big/ \tau_{\mathrm{U},ph-ph,m_p} \\
\Phi_{m_p,i,j}^{eq,t+1} &= \alpha_{ph-e}\phi_{m_p,i,j}^{eq,t+1}\left(\tilde{T}_e\right) + \alpha_{ph-ph}\phi_{m_p,i,j}^{eq,t+1}\left(\tilde{T}_{ph}\right)
\end{aligned} \qquad \text{(B7)}
$$

The pseudo-temperature of phonons and electrons are thus calculated based on the numerical solution of Eqs. (30) and (31) by dichotomy or Newton's method:

$$\sum_p \frac{\omega_{\max,p}}{2} \sum_{m_p=1}^{N_{ph,p}} \frac{\phi_{m_p}^{eq}\left(\tilde{T}_{ph,i,j}^{t+1}\right)}{\left(v_{ph}\tau_{U,ph-ph}\right)_{m_p}} w_{m_p} = \pi \sum_p \frac{\omega_{\max,p}}{2} \sum_{m_p=1}^{N_{ph,p}} \sum_{\theta=1}^{N_\theta} \sum_{\varphi=1}^{N_\varphi} \frac{\phi_{m_p,i,j}^{\theta,\varphi,t+1}}{\left(v_{ph}\tau_{U,ph-ph}\right)_{m_p}} w_\varphi w_\theta w_{m_p} \qquad ,$$

(B8)

$$\varepsilon_{\text{HFW}} \sum_{n=1}^{N_e} \frac{I_n^{eq}\left(\tilde{T}_{e,i,j}^{t+1}\right)}{\left(v_e\tau_{e-ph}\right)_n} w_n + \sum_p \frac{\omega_{\max,p}}{2} \sum_{m_p=1}^{N_{ph,p}} \frac{\phi_{m_p}^{eq}\left(\tilde{T}_{e,i,j}^{t+1}\right)}{\left(v_{ph}\tau_{ph-e}\right)_{m_p}} w_{m_p}$$

$$= \pi \varepsilon_{\text{HFW}} \sum_{n=1}^{N_e} \sum_{\theta=1}^{N_\theta} \sum_{\varphi=1}^{N_\varphi} \frac{I_{n,i,j}^{\theta,\varphi,t+1}}{\left(v_e\tau_{e-ph}\right)_n} w_\varphi w_\theta w_n + \pi \sum_p \frac{\omega_{\max,p}}{2} \sum_{m_p=1}^{N_{ph,p}} \sum_{\theta=1}^{N_\theta} \sum_{\varphi=1}^{N_\varphi} \frac{\phi_{m_p,i,j}^{\theta,\varphi,t+1}}{\left(v_{ph}\tau_{ph-e}\right)_{m_p}} w_\varphi w_\theta w_{m_p}$$

, (B9)

where $w_\varphi$, $w_\theta$, $w_n$ and $w_{m_p}$ are the corresponding weight coefficients with $\varepsilon_{\text{HFW}}$ the half-width of the Fermi window. Similarly, the macroscopic variables including local electron, phonon temperature and e-ph coupling constant are computed based on formulas (32)-(34):

$$\varepsilon_{\text{HFW}} \sum_{n=1}^{N_e} \frac{I_n^{eq}\left(T_{e,i,j}^{t+1}\right)}{v_{e,n}} w_n = \pi \varepsilon_{\text{HFW}} \sum_{n=1}^{N_e} \sum_{\theta=1}^{N_\theta} \sum_{\varphi=1}^{N_\varphi} \frac{I_{n,i,j}^{\theta,\varphi,t+1}}{v_{e,n}} w_\varphi w_\theta w_n \,, \tag{B10}$$

$$\sum_p \frac{\omega_{\max,p}}{2} \sum_{m_p=1}^{N_{ph,p}} \frac{\phi_{m_p}^{eq}\left(T_{ph,i,j}^{t+1}\right)}{v_{ph,m_p}} w_{m_p} = \pi \sum_p \frac{\omega_{\max,p}}{2} \sum_{m_p=1}^{N_{ph,p}} \sum_{\theta=1}^{N_\theta} \sum_{\varphi=1}^{N_\varphi} \frac{\phi_{m_p,i,j}^{\theta,\varphi,t+1}}{v_{ph,m_p}} w_\varphi w_\theta w_{m_p} \,, \tag{B11}$$

$$G_{i,j}^{t+1} = \frac{\sum_p \frac{\omega_{\max,p}}{2} \sum_{m_p=1}^{N_{ph,p}} \frac{\phi_{m_p}^{eq}\left(\tilde{T}_{e,i,j}^{t+1}\right)}{\left(v_{ph}\tau_{ph-e}\right)_{m_p}} w_{m_p} - \pi \sum_p \frac{\omega_{\max,p}}{2} \sum_{m_p=1}^{N_{ph,p}} \sum_{\theta=1}^{N_\theta} \sum_{\varphi=1}^{N_\varphi} \frac{\phi_{m_p,i,j}^{\theta,\varphi,t+1}}{\left(v_{ph}\tau_{ph-e}\right)_{m_p}} w_\varphi w_\theta w_{m_p}}{\tilde{T}_{e,i,j}^{t+1} - \tilde{T}_{ph,i,j}^{t+1}} \,.$$

(B12)

Besides, the isothermal boundary for $x$ direction and periodic boundary for $y$ direction are shown respectively:

$$\begin{aligned} &I_{n,x=0,j}^{u_\theta>0,\varphi,t+1} = I_n^{eq}\left(T_h\right), I_{n,x=L,j}^{u_\theta<0,\varphi,t+1} = I_n^{eq}\left(T_c\right) \\ &\phi_{m_p,x=0,j}^{u_\theta>0,\varphi,t+1} = \phi_{m_p}^{eq}\left(T_h\right), \phi_{m_p,x=L,j}^{u_\theta<0,\varphi,t+1} = \phi_{m_p}^{eq}\left(T_c\right) \end{aligned}, \tag{B13}$$

$$I_{n,i,y=0}^{\theta,\eta>0,t+1} = I_{n,i,y=H}^{\theta,\eta>0,t+1}, \ I_{n,i,y=H}^{\theta,\eta<0,t+1} = I_{n,i,y=0}^{\theta,\eta<0,t+1}$$
$$\phi_{m_\mathrm{p},i,y=0}^{\theta,\eta>0,t+1} = \phi_{m_\mathrm{p},i,y=H}^{\theta,\eta>0,t+1}, \ \phi_{m_\mathrm{p},i,y=H}^{\theta,\eta<0,t+1} = \phi_{m_\mathrm{p},i,y=0}^{\theta,\eta<0,t+1},$$

(B14)

whereas other types of boundaries can be directly referred from our previous work [40].

Therefore, the computational procedures of DOM scheme for the coupled electron and phonon BTEs is as followed:

(a) Initialize the electron, phonon intensity and pseudo-equilibrium intensity field at the moment $t$+1 by that of the last time step;

(b) Implement the boundary conditions;

(c) Update the electron and phonon intensity field by Eqs. (B5) and (B6);

(d) Calculate the electron and phonon pseudo-temperature based on Eqs. (B8) and (B9), and the corresponding pseudo-equilibrium intensity;

(e) Compare the pseudo-equilibrium intensity with that of the last iteration; if the convergence criterion is not satisfied, go back to step (b) until it is satisfied;

(f) Compute the macroscopic variables and then go to the next time step and loop from step (a) until finishing the prescribed time steps or reaching the steady state.